\documentclass[review]{elsarticle}

\usepackage{wrapfig}
\usepackage{graphicx}
\usepackage{subfig}
\usepackage[cmex10]{amsmath} 
\usepackage{amssymb}
\usepackage{amsmath}
\usepackage{accents}
\usepackage[nottoc]{tocbibind}
\usepackage{mathtools}
\usepackage{amsthm}
\usepackage{amsfonts}

\usepackage{steinmetz}

\usepackage{color}

\usepackage[framed,numbered]{mcode}

\usepackage[hidelinks,unicode]{hyperref}

\newtheorem{theorem}{Theorem}
\newtheorem{proposition}{Proposition}

\newtheorem{corollary}{Corollary}
\newtheorem{remark}{Remark}
\newtheorem{example}{Example}

\newcommand{\SNR}{\operatorname{SNR}}
\newcommand{\be}{\vspace*{-0.035cm}\begin{eqnarray}}
\newcommand{\ee}{\end{eqnarray}\vspace*{-0.035cm}}
\newcommand{\beIEEE}{\vspace*{-0.02cm}\begin{IEEEeqnarray}{rCl}}
\newcommand{\eeIEEE}{\end{IEEEeqnarray}\vspace*{-0.02cm}}

\usepackage[linesnumbered,ruled]{algorithm2e}

\newcommand{\CN}{\mathcal{CN}}
\newcommand{\N}{\mathcal{N}}
\newcommand{\MSE}{\operatorname{MSE}}
\newcommand{\E}{\mathbb{E}}

\newcommand{\MCRB}{\operatorname{MCRLB}}
\newcommand{\HCRB}{\operatorname{HCRB}}

\newcommand{\BB}{\operatorname{BB}}
\newcommand{\CD}{\boldsymbol{\operatorname{C}}}

\renewcommand{\C}{\mathcal{C}}
\newcommand{\I}{\mathcal{I}}

\renewcommand{\L}{\mathcal{L}}

\newcommand{\myfrac}[3][0pt]{\genfrac{}{}{}{}{\raisebox{#1}{$#2$}}{\raisebox{-#1}{$#3$}}}  

\renewcommand{\d}{\;\mathrm{d}{}}
\renewcommand{\t}{\mathrm{T}{}}
\newcommand{\h}{\mathrm{H}{}}
\renewcommand{\H}{\mathcal{H}}
\newcommand{\tr}{\mathrm{tr}}

\journal{Signal Processing}

\newcommand{\insertfigure}[3]%
{%
	\vskip10pt
	\begin{figure}
		\centering
			\includegraphics[width=0.4\textwidth]{#1}
			\caption{#2}
			\label{#3}
	\end{figure}
}

\begin{document}
\interdisplaylinepenalty=0

\begin{frontmatter}



\title{An Approximate MSE Expression for Maximum Likelihood and Other Implicitly Defined Estimators of Non-Random Parameters \newline (extended version)}


\author[METU]{Erdal Mehmetcik}
\ead{mehmetcik.erdal@metu.edu.tr}

\author[METU]{Umut Orguner}
\ead{umut@metu.edu.tr}

\author[METU]{{\c{C}}a{\u{g}}atay Candan \corref{cor1}}
\ead{ccandan@metu.edu.tr}

\address[METU]{Department of Electrical and Electronics Engineering, Middle East Technical University (METU), 06800, Ankara, Turkey.}

\cortext[cor1]{Corresponding author}

\begin{abstract}
An approximate mean square error (MSE) expression for the performance analysis of implicitly defined estimators of non-random parameters is proposed. An implicitly defined estimator (IDE) declares the minimizer/maximizer of a selected cost/reward function as the parameter estimate. The maximum likelihood (ML) and the least squares estimators are among the well known examples of this class. In this paper, an exact MSE expression for implicitly defined estimators with a symmetric and unimodal objective function is given. It is shown that the expression reduces to the Cramer-Rao lower bound (CRLB) and misspecified CRLB in the large sample size regime for ML and misspecified ML estimation, respectively. The expression is shown to yield the Ziv-Zakai bound (without the valley filling function) for the maximum a posteriori (MAP) estimator when it is used in a Bayesian setting, that is, when an a-priori distribution is assigned to the unknown parameter. In addition, extension of the suggested expression to the case of nuisance parameters is studied and some approximations are given to ease the computations for this case. Numerical results indicate that the suggested MSE expression not only predicts the estimator performance in the asymptotic region; but it is also applicable for the threshold region analysis, even for IDEs whose objective functions do not satisfy the symmetry and unimodality assumptions. Advantages of the suggested MSE expression are its conceptual simplicity and its relatively straightforward numerical calculation due to the reduction of the estimation problem to a binary hypothesis testing problem, similar to the usage of Ziv-Zakai bounds in random parameter estimation problems.
\end{abstract}




\end{frontmatter}


\setcounter{page}{1}
\section{Introduction}
The topic of parameter estimation can be divided into two classes, namely the estimation of random and non-random (deterministic) parameters. The random parameter estimation (Bayesian estimation) assumes that the parameter of interest is a random variable with an a-priori distribution and the observations on a realization of the unknown parameter are obtained according to a known probabilistic mapping. Under this setting, the optimal estimator that minimizes the risk, say mean square error (MSE) or mean absolute error, is  a functional of the posterior density of the parameter. For instance, the optimal estimator minimizing the MSE is the mean value of the parameter with respect to the posterior density~\cite{Kay_EstimationTheory}. In general, the posterior density calculation is the key step for the Bayesian formulation. Unfortunately, a closed form expression for the posterior density (and its moments) which does not involve integration, differentiation and limit operations is rarely available. In many problems, one has to resort to the Monte Carlo methods or approximate inference techniques for an inexact realization of the optimal Bayesian estimator. In such problems, the estimator success is typically evaluated by comparisons with the performance bounds.
Bayesian performance  bounds have a vast literature \cite{VanTrees_BayesianBounds}. Typically, these bounds do not impose any constraints on the estimator. For instance the Bayesian Cramer Rao lower bound (CRLB) \cite{VanTrees_BayesianBounds, vanTreesDetection}, Weiss Weinstein bound (WWB) \cite{WeissWeinsteinBound}, Bayesian Bhattacharya bound \cite{vanTreesDetection} are derived using the covariance inequality principle (hence, sometimes referred as covariance bounds) and applicable in general to any type of estimators. Another main class is the Ziv-Zakai bound (ZZB) \cite{ZivZakaiBound, BelliniTartara1974, BellSEVT:1997} type bounds which are derived by converting the estimation problem into a binary detection problem. Bayesian CRLB is one of most fundamental bounds and provides the achievable MSE in the asymptotic region which is the high signal-to-noise ratio (SNR) region. However, it suffers from the threshold effect \cite{ZivZakaiBound, BelliniTartara1974, BellSEVT:1997}, meaning that it provides unachievable (optimistic) lower bounds at medium or low SNR values. ZZB and WWB are among the tightest Bayesian bounds in all regions of operation \cite{VanTrees_BayesianBounds, BellSEVT:1997}. Bayesian bounds continue to be an active research area. Recently, Bayesian bounds for estimating periodic parameters (e.g., phase)  have been developed \cite{2016_NitzanRouttenbergTabrikian_ClassOfBayesianCyclicBounds, 2020_XuColeman_MinimaxLowerBoundsForCircularLocalization, 2022_Tabrikian_BPCRB}.

Non-random parameter estimation involves some challenges unique to this setting. In this setting, an estimator can be improved for a specific value of the parameter at the expense of performance for other parameter values \cite{Kay_EstimationTheory,GallagerBook}. Since there is no a-priori distribution associated to the parameter of interest, it is not possible to balance the performance gains and losses for different parameter values as in the Bayesian setting. For example, the estimator ignoring the measurements and producing a constant value, say $\alpha$, as the estimate has no error if the unknown parameter is indeed $\alpha$; but, suffers from performance losses at all other parameter values. The development of lower bounds for the non-random parameter estimation also suffers from similar inherent admissibility problems. To overcome these problems, the estimators in this setting are typically restricted to the class of unbiased estimators and examined under the title of  minimum variance unbiased estimators \cite{Kay_EstimationTheory}.

The performance bounds for the non-random parameter estimation are also developed for a specific class of estimators. For example CRLB (for non-random parameters), Hammersley-Chapman-Robbins Bound (HCRB) \cite{ChapmanRobbinsBound}, Barankin Bound \cite{BarankinBound,BarankinBound1971} require the estimator to be unbiased in an open neighborhood of a point, over a set of two-points and over a set of many-points, respectively (also see \cite{ForsterLarzabal2002}). In \cite{TodrosTabrikian2010}, a general bound form for unbiased estimators is given and it is shown that CRLB, HCRB and BB can be derived by a proper choice of the kernel function of their integral transform. Note that the unbiasedness condition may not be practical or may be difficult to satisfy, especially for the parameters with a finite support;  since the estimation error approaches a one-sided distribution at the edges of the parameter space in such cases \cite{ZivZakaiBound}. Although it has been shown that for some problems with periodic parameters
 \cite{2013_Routtenberg_NonBayesianPeriodicCRB, 2014_Tabrikian_CyclicBarankin,2016_RouttenbergTabrikian_CyclicCRBTypeBounds}, the problem with the one-sided error distribution at the edges may vanish; uniformly unbiased estimators do not exist in these cases either \cite{2015_TodrosTabrikian_barankinLimitations}. Furthermore, the unbiasedness condition may not even be desirable in some problems. It is known that there exist realizable biased-estimators for some problems whose MSE is lower than the Cramer-Rao bound for unbiased estimators \cite{stoica1990biasedest-CRB}. Perhaps, the most important aspect of unbiasedness condition is in relation with the maximum likelihood estimator. It is well known that the maximum likelihood estimator is unbiased and efficient in the large sample size regime, under fairly general conditions, providing a basis for the theoretical and practical adoption of the unbiasedness condition \cite{White1981,WassermanAllofStatistics}.

The main problem considered in this paper is the performance prediction of implicitly defined estimators (IDEs) of non-random parameters. IDEs are estimators which produce an estimate by maximizing an objective function of the measurements over the parameter set under consideration. The maximum likelihood (ML) estimator, least squares estimators are some well known examples.

In this study, we present an approximate MSE expression for IDEs of  non-random parameters that
\begin{itemize}
\item gives the true MSE when the objective function of the IDE is symmetric and unimodal,
    \item reduces to the CRLB in the large sample size regime for ML estimation,
    \item reduces to the misspecified CRLB (MCRLB)~\cite{FortunatiGGR2017,Richmond2015_MCRB} in the large sample size regime for misspecified ML estimation~\cite{Richmond2015_MCRB, white82},
    \item reduces to the ZZB when an a-priori distribution is assigned to the  parameter of interest for maximum a posteriori (MAP) estimation.
\end{itemize}

There are already some approximate MSE expressions available in the literature for IDEs of non-random parameters. For instance, \cite{Fessler1996} provides  formulas for MSE and bias of IDEs using Taylor series expansion of the cost function along with some approximate expressions for certain expectations and derivatives. The study in \cite{SimpleFormulasForBiasAndMSE} also uses Taylor expansion approach and derives different approximations in the scalar parameter case. Both of these approaches are based on the Taylor series expansion around the true parameter value and provide simple MSE expressions; but, do not take into account the gross errors which becomes the significant factor as the estimator nonlinearity increases and/or SNR is decreased below the threshold SNR.

The majority of work on the estimator performance prediction focus on the performance of the ML estimator. The ML estimator is known to be asymptotically efficient (performance approaching CRLB at large sample size) under some regularity conditions \cite{vanTreesDetection, White1981, WassermanAllofStatistics}.  The method of interval errors (MIE) is a celebrated method that was proposed by Van Trees \cite{vanTreesDetection}  to assess the performance of the ML estimator in the threshold region.  This method depends on a careful selection of intervals in the parameter space and the calculation of their probabilities. Different approximations have been proposed to approximate the probabilities~\cite{Athley_thresholdSNR_2005, Richmond2006, SinanGezici_StatisticsOfMLE}. The MSE expression proposed in the present study can be interpreted as a more principled version of the method of interval errors where the need for the interval selection and the gross error probability calculation or approximation is not required.

Notation: Throughout the paper lower and uppercase letters denote scalars, e.g., $a$, $A$. Bold lowercase letters denote vectors, e.g., $\mathbf{a}$. Bold uppercase letters denote matrices, e.g., $\mathbf{A}$. The $i$th element of the vector $\mathbf{a}$ is denoted by $[\mathbf{a}]_i$. The $i,j$th element of the matrix $\mathbf{A}$ is denoted by $[\mathbf{A}]_{i,j}$. $\Re\{\cdot\}$ denotes the real part of the complex argument.
\section{Problem Definition}
We consider the estimation of the non-random real-valued vector $\boldsymbol{\theta}\triangleq[\theta_1\,\,\theta_2\,\cdots\,\,\theta_J]^\t$ from the  measurements $\mathbf{x}\triangleq[x_0\,\,x_1\,\cdots\,\,x_{N-1}]^\t\in\mathbb{C}^N$ distributed according to $f(\mathbf{x};\boldsymbol{\bar\theta})$ where $\boldsymbol{\bar\theta}\triangleq[\bar\theta_1\,\,\bar\theta_2\,\cdots\,\,\bar\theta_J]^\t$ denotes the true value of $\boldsymbol{\theta}$. An implicitly defined estimator (IDE) generates an estimate $\boldsymbol{\hat\theta}\triangleq[\hat\theta_1\,\,\hat\theta_2\,\cdots\,\,\hat\theta_J]^\t$ by maximizing an objective function $\L(\cdot,\cdot)$ of the measurements and the parameters as shown below: \begin{align}
    \boldsymbol{\hat\theta}\triangleq\arg\max_{\boldsymbol{\theta}} \L(\mathbf{x};\boldsymbol{\theta}).
\label{eqn:IDEdefinition}
\end{align}
The most well-known example of IDEs is the ML estimator where the objective function $\L(\cdot,\cdot)$ is the likelihood function $f(\mathbf{x};\boldsymbol{\theta})$. Other examples of IDEs are M-estimators and (nonlinear) least square estimators. We see that the estimate $\boldsymbol{\hat\theta}$ given by \eqref{eqn:IDEdefinition} is determined by the measurements implicitly, hence the name \emph{implicitly defined estimator}.

In this study we are interested in the performance of IDEs and we give an expression for the (diagonal elements of the) MSE matrix of the estimate $\boldsymbol{\hat\theta}$ which is defined as
\begin{align}
\MSE(\boldsymbol{\bar\theta})\triangleq&\E\big[(\boldsymbol{\hat\theta}-\boldsymbol{\bar\theta})(\boldsymbol{\hat\theta}-\boldsymbol{\bar\theta})^\mathrm{T}\big],
\end{align}
Here it should be mentioned that the methodology presented in the current work can be straightforwardly extended to the other moments of the estimation error $\boldsymbol{\hat\theta}-\boldsymbol{\bar\theta}$.

Except for few cases like ML estimation for Gaussian likelihoods with linear models, the estimate $\boldsymbol{\hat\theta}$ in~\eqref{eqn:IDEdefinition} cannot be analytically expressed in terms of the measurements $\mathbf{x}$, i.e., one cannot find a closed form expression for the function $\mathbf{h}(\cdot)$ such that $\boldsymbol{\hat\theta}=\mathbf{h}(\mathbf{x})$. As a consequence the determination, evaluation and comparison of performance (say, in terms of MSE) of an IDE usually involves extensive Monte Carlo studies and/or problem specific  approximations.
In this work we first give an MSE expression which is exact for an IDE of a scalar parameter whose objective function is both symmetric (around the estimate) and unimodal in Section \ref{Sec3SymUniModal}. Since the symmetry and unimodality conditions are typically satisfied by the objective functions of IDEs in the asymptotic or small error region, as further examined in Section \ref{Sec4ReltoOthers}; we suggest to use the MSE expression to study the performance of IDEs
in the small error and threshold regions. We refrain from calling the suggested MSE expression as a bound due to the lack of performance guarantees in the  non-asymptotic regions. The suggested expression can be considered to be in the same league with the MIE \cite{vanTreesDetection} which lacks a performance guarantee in all regions including the asymptotic region. Such expressions are also called approximate bounds in some studies \cite{SinanGezici_StatisticsOfMLE,Mallat_Gezici_MLEpaper}. Yet, our main goal in this study is to develop an MSE expression similar to ZZB, which is known to be a tight random parameter estimation bound in the threshold and asymptotic regions, for non-random parameters.

\section{Case of a Scalar Parameter with Symmetric and Unimodal Objective Functions}
\label{Sec3SymUniModal}
In this section we are going to restrict ourselves to a scalar unknown parameter $\theta\in\mathbb{R}$ (i.e., $J=1$) and provide a predicted MSE expression which is equal to the true MSE for an IDE whose objective function satisfies symmetry and unimodality assumptions. Our main results are given in the following theorem and its corollary.
\begin{figure}
		\begin{center}
		\includegraphics[width=0.70\columnwidth]{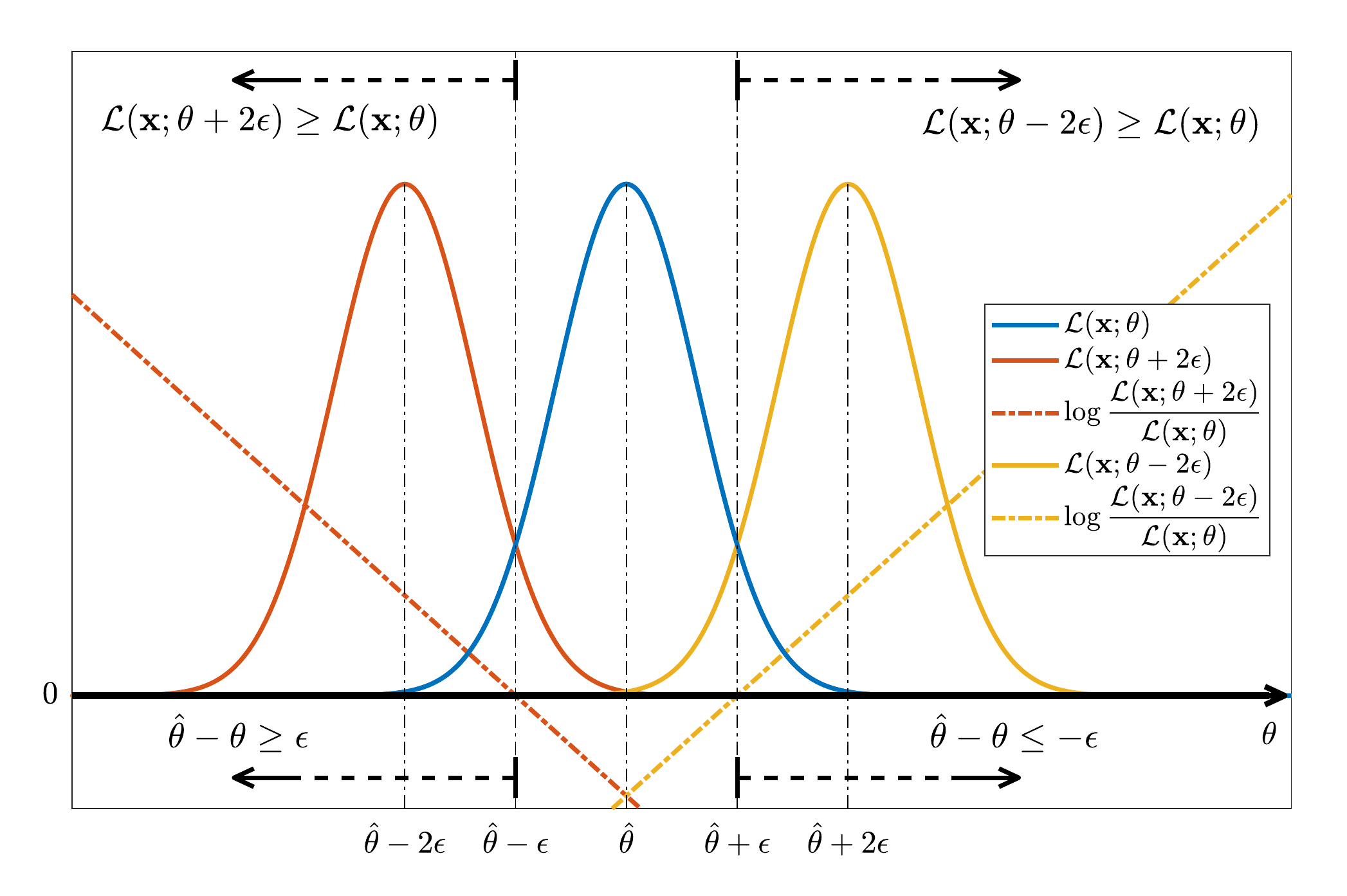}
		\caption{Illustration of the fact that the events $(\hat\theta-\theta)\ge\epsilon$ and $(\hat\theta-\theta)\le -\epsilon$ are equivalent to the events $\L(\mathbf{x};\theta+2\epsilon) \ge \L(\mathbf{x};\theta)$ and $\L(\mathbf{x};\theta-2\epsilon) \ge \L(\mathbf{x};\theta)$, respectively, when the
assumptions of Theorem~\ref{thm:IDEV} hold.}
		\label{fig:TheoremIllustration}
		\end{center}
\end{figure}
\begin{theorem}\label{thm:IDEV} Consider the IDE given as
\begin{align}
   \hat\theta\triangleq\arg\max_{\theta} \L(\mathbf{x};\theta).\label{IDEinTheorem}
\end{align}
Let the objective function $\L(\mathbf{x};\cdot)$ satisfy the following conditions for all $\mathbf{x}\in\mathbb{C}^N$.
\begin{enumerate}
\item $\L(\mathbf{x};\hat{\theta}+h)=\L(\mathbf{x};\hat{\theta}-h)$ for all $h\in\mathbb{R}$, i.e., the objective function is symmetric around its peak.
\item $\L(\mathbf{x};\theta)$ is strictly-increasing (strictly-decreasing) for $\theta<\hat{\theta}$ ($\theta>\hat{\theta}$).
\end{enumerate}
Define the true estimator statistic $V_{\hat{\theta}}(\theta)$ as
\begin{align}
    V_{\hat{\theta}}(\theta)\triangleq E[(\hat{\theta}-\theta)^2],\label{eqn:DefinitionofV}
\end{align}
where $\theta$ is an arbitrary fixed parameter value. Then,
\begin{align}
V_{\hat{\theta}}(\theta)&=\widehat{V}_{\hat{\theta}}(\theta),\label{eqn:VsAreEqual}
\end{align}
where the predicted statistic $\widehat{V}_{\hat{\theta}}(\theta)$ is defined as
\begin{align}
\Aboxed{
\widehat{V}_{\hat{\theta}}(\theta)&\triangleq 2\int_{-\infty}^\infty
\! |\epsilon|P\left(\L(\mathbf{x};\theta+2\epsilon)\ge \L(\mathbf{x};\theta)\right)
\!\!\d\epsilon.
\label{eqn:VHATdefinition}
}
\end{align}
\end{theorem}
\textbf{Proof:} A proof is presented in~\ref{app:proofofTheoremIDEbiasandMSE}. The main idea of the proof is to show that the events $(\hat\theta-\theta)\ge \epsilon$ and  $(\hat\theta-\theta)\le-\epsilon$ are equivalent to the events $\L(\mathbf{x};\theta+2\epsilon)\ge \L(\mathbf{x};\theta)$ and $\L(\mathbf{x};\theta-2\epsilon)\ge\L(\mathbf{x};\theta)$, respectively, when the
assumptions of the theorem hold. An illustration of this equivalence is given in Figure~\ref{fig:TheoremIllustration}.\hfill$\blacksquare$

The following remark applies Theorem~\ref{thm:IDEV} to find the true MSE of the estimator $\hat{\theta}$.

\begin{remark}[MSE of IDE]\label{cor:MSE}  The true MSE of the IDE $\hat{\theta}$ in Theorem~\ref{thm:IDEV} is given as
\begin{align}
\MSE(\bar\theta)&=\widehat{\MSE}(\bar\theta)
\label{eqn:MSEequalityforIDE}
\end{align}
where the predicted MSE, denoted as $\widehat{\MSE}(\bar{\theta})$, is defined as
\begin{align}
\Aboxed{\widehat{\MSE}(\bar\theta)&\triangleq 2\int_{-\infty}^\infty
\! |\epsilon|P\left(\L(\mathbf{x};\bar{\theta}+2\epsilon)\ge \L(\mathbf{x};\bar{\theta})\right)
\!\!\d\epsilon.}
\label{eqn:MSEHATdefinition}
\end{align}

\end{remark}
\textbf{Proof:} The proof is trivial by realizing that $\MSE(\bar\theta)=V_{\hat{\theta}}(\bar{\theta})$ and $\widehat{\MSE}(\bar\theta)=\widehat{V}_{\hat{\theta}}(\bar{\theta})$.
\hfill$\blacksquare$

In the special case of a parameter $\theta$ with finite support, e.g., $\theta \in [\theta_{\min},\,\theta_{\max}]$, the estimation error $\hat\theta-\bar\theta$ is restricted to the interval $[\theta_{\min}-\bar\theta,\,\theta_{\max}-\bar\theta]$ and the predicted MSE becomes
\begin{align}
    \widehat{\MSE}(\bar{\theta})=\,2\int_{\frac{\theta_{\min}-\bar\theta}{2}}^{\frac{\theta_{\max}-\bar\theta}{2}}|\epsilon|P\left( \L(\mathbf{x};\bar{\theta}+2\epsilon)\ge\L(\mathbf{x};\bar{\theta})\right)\d\epsilon.\label{eqn:finiteSupport}
\end{align}

We can interpret the MSE expression~\eqref{eqn:MSEHATdefinition} intuitively as follows. When the probability $P\left(\L(\mathbf{x};\bar{\theta}+2\epsilon)\ge \L(\mathbf{x};\bar{\theta})\right)$ is large for large values of $|\epsilon|$, then it is probable for the IDE $\hat\theta$ to make gross errors, resulting in a large MSE. On the other hand, if this probability is small for large values of $|\epsilon|$, the contribution of gross errors in the MSE becomes negligible, resulting in a small MSE. Consequently, IDEs with a small MSE would have the probability $P\left(\L(\mathbf{x};\bar{\theta}+2\epsilon)\ge \L(\mathbf{x};\bar{\theta})\right)$ (thought of as a function of $\epsilon$) highly concentrated in a small neighborhood of $\epsilon=0$ and \emph{quickly} vanishing elsewhere. More specifically, a sufficient and necessary condition for existence of the integral in the MSE expression~\eqref{eqn:MSEHATdefinition} is $P\left( \L(\mathbf{x};\bar{\theta}+2\epsilon)\ge\L(\mathbf{x};\bar{\theta})\right)= o(1/|\epsilon|^2)$, i.e., the probability $P\left( \L(\mathbf{x};\bar{\theta}+2\epsilon)\ge\L(\mathbf{x};\bar{\theta})\right)$ decaying strictly faster than $1/|\epsilon|^2$ as $|\epsilon|\rightarrow\infty$. The integral in the MSE expression~\eqref{eqn:finiteSupport}, on the other hand, always exists.

We can put the expression $\widehat{\MSE}(\bar\theta)$ to a test by considering the  optimal but infeasible estimator $\hat\theta=\bar{\theta}$. This estimator can be formulated as an IDE using the objective function $\L(\mathbf{x};\theta)\triangleq-(\theta-\bar{\theta})^2$. Since the objective function $\L(\mathbf{x};\cdot)$ does not depend on the measurements $\mathbf{x}$, we see that the probability of the deterministic event $\L(\mathbf{x};\bar{\theta}+2\epsilon)\ge \L(\mathbf{x};\bar{\theta})$ is given as
\begin{align}
    P\left(\L(\mathbf{x};\bar{\theta}+2\epsilon)\ge\L(\mathbf{x};\bar{\theta}) \right)=\begin{cases}1,&\epsilon=0\\0,&\text{otherwise}\end{cases}.
\label{probsanity} \end{align}
When \eqref{probsanity} is substituted into~\eqref{eqn:MSEHATdefinition}, we have $\widehat{\MSE}(\bar{\theta})=0$, which is the true MSE. Similarly, for the feasible (but biased) version of this estimator $\hat{\theta}=\theta_0$, where $\theta_0 \in \mathbb{R}$, the objective function is $\L(\mathbf{x};\theta)\triangleq-(\theta-\theta_0)^2$ and the corresponding probability becomes
\begin{equation}
    P\left(\L(\mathbf{x};\bar{\theta}+2\epsilon)\ge\L(\mathbf{x};\bar{\theta}) \right)
      = \begin{cases}1,& -(\bar{\theta}+2\epsilon-\theta_0)^2 \ge -(\bar{\theta}-\theta_0)^2 \\0,&\text{otherwise} \end{cases} =\begin{cases}1,&0 \ge \epsilon \ge -\bar{\theta}+\theta_0 \\1,& -\bar{\theta}+\theta_0 \ge \epsilon \ge 0 \\0,&\text{otherwise} \end{cases},
    \label{probsanity_biased}
\end{equation}
which yields $\widehat{\MSE}(\bar\theta) = (\bar{\theta}-\theta_0)^2$ when substituted into~\eqref{eqn:MSEHATdefinition} for both $\theta_0 \ge \bar{\theta}$ and $\theta_0 \le \bar{\theta}$. This also is the true MSE.
The following corollary applies the result in Theorem~\ref{thm:IDEV} to ML estimation.
\begin{corollary}[MSE of ML Estimator]\label{cor:MLcase} If the likelihood function $f(\mathbf{x};\cdot)$ satisfies the conditions in Theorem~\ref{thm:IDEV}, then the true MSE of the ML estimate $\hat{\theta}$ is given as
\begin{align}
\MSE_{\mathrm{ML}}(\bar{\theta})=\widehat{\MSE}_{\mathrm{ML}}(\bar{\theta})
\end{align}
where the predicted MSE, denoted as $\widehat{\MSE}_{\mathrm{ML}}(\bar{\theta})$, is defined as
\begin{align}
\Aboxed{
\widehat{\MSE}_{\mathrm{ML}}(\bar{\theta})\triangleq&\,2\int_{-\infty}^\infty|\epsilon|P\bigg( \myfrac[1.5pt]{f(\mathbf{x};\bar{\theta}+2\epsilon)}{f(\mathbf{x};\bar{\theta})}\ge 1\bigg)\d\epsilon.
}
\label{eqn:MSEHATMLdefinition}
\end{align}
\end{corollary}
\textbf{Proof:} Since we have $\L(\mathbf{x};\theta)\triangleq f(\mathbf{x};\theta)$, we can write
\begin{subequations}
\begin{align}
   P\left(\L(\mathbf{x};\bar{\theta}+2\epsilon)\ge\L(\mathbf{x};\bar{\theta})\right) &= P\left(f(\mathbf{x};\bar{\theta}+2\epsilon)\ge f(\mathbf{x};\bar{\theta})\right)=\int I\left(f(\mathbf{x};\bar{\theta}+2\epsilon)\ge f(\mathbf{x};\bar{\theta})\right)f(\mathbf{x};\bar{\theta})\d\mathbf{x}\\
   &\hspace{-4cm}=\int_{f(\mathbf{x};\bar{\theta})\neq 0} I\left(f(\mathbf{x};\bar{\theta}+2\epsilon)\ge f(\mathbf{x};\bar{\theta})\right)f(\mathbf{x};\bar{\theta})\d\mathbf{x}=\int_{f(\mathbf{x};\bar{\theta})\neq 0} I\bigg(\myfrac[1.5pt]{f(\mathbf{x};\bar{\theta}+2\epsilon)}{f(\mathbf{x};\bar{\theta})}\ge 1\bigg)f(\mathbf{x};\bar{\theta})\d\mathbf{x}\\
   &\hspace{-4cm}=P\bigg(\myfrac[1.5pt]{f(\mathbf{x};\bar{\theta}+2\epsilon)}{f(\mathbf{x};\bar{\theta})}\ge 1\bigg),
\end{align}
\end{subequations}
where $I(\cdot)$ denotes the indicator function for event arguments. Substituting the last probability into the integral in~\eqref{eqn:MSEHATdefinition} completes the proof.\hfill$\blacksquare$

Note that the expression~\eqref{eqn:MSEHATMLdefinition} connects the MSE of the ML estimate to the error probability of a likelihood ratio test. This connection between estimation and detection theory is further explored in Section~\ref{sec:relationtoZZB} in relation with  the ZZB.

Theorem~\ref{thm:IDEV} and its corollary provide compact expressions to evaluate the MSE of an implicitly defined estimator exactly even when there is no explicit analytical expression connecting the estimate $\hat{\theta}$  to the measurements $\mathbf{x}$. However, it has some limitations imposed by the assumptions required for its validity. In fact, almost all practical estimation problems violate one of the assumptions of symmetry, unimodality and infinite support (of the parameter $\theta$). For these problems it is certainly possible to have $\MSE(\bar\theta)\neq\widehat{\MSE}(\bar\theta)$. Hence, in general the proposed expressions in~\eqref{eqn:MSEHATdefinition}  and~\eqref{eqn:MSEHATMLdefinition} can only serve as
approximate MSE performance prediction tools. Keeping this fact in mind, we show several relations between the suggested MSE expression and well-known bounds in Section~\ref{Sec4ReltoOthers}.

\section{Relationship to Performance Bounds}
\label{Sec4ReltoOthers}
In this section we  present the  relationship of the suggested MSE expression~\eqref{eqn:MSEHATMLdefinition} to some well-known performance bounds.

\subsection{Relationship to CRLB}\label{sec:relationtoCRLB}
In this section, we consider the ML estimation for a scalar parameter $\theta\in\mathbb{R}$. In order to use the large sample asymptotic results for the ML estimate $\hat\theta$, we are going to assume that the elements $x_n$, $n=0,\ldots,N-1$, of the measurement vector $\mathbf{x}$ are independent and identically distributed as $x_n\sim f(x_n;\bar{\theta})$~\footnote{In order not to incorporate additional notation, we will keep here the individual measurements $x_n\in\mathbb{C}$ as scalars but the same results can be obtained for the case when $x_n$ is a vector.}. The likelihood for the measurement vector $\mathbf{x}$ is then given as
\begin{align}
    f(\mathbf{x};\theta)=\prod_{n=0}^{N-1} f(x_n;\theta).\label{eqn:iid-likelihood}
\end{align}
The relationship of the suggested MSE expression $\widehat{\MSE}_{\mathrm{ML}}(\bar{\theta})$ to CRLB is given in the following proposition.

\begin{proposition}\label{prop:CaseofCRLB}
\begin{itemize} Assume that
    \item[\textbf{A0}] The parameter $\theta$ has finite support, i.e., $\theta\in[\theta_{\mathrm{min}},\theta_{\mathrm{max}}]$, and the true parameter value $\bar\theta$ satisfies $\bar\theta\in(\theta_{\mathrm{min}},\theta_{\mathrm{max}})$.
    \item[\textbf{A1}] The first three derivatives of $\ln f(x;\theta)$ with respect to $\theta$ exist for all $\theta$ and are continuous with respect to $\theta$.
    \item[\textbf{A2}] For every $\theta$, the functions $|\frac{\partial^i}{\partial\theta^i}\ln f(x;\theta)|$, $i=0,1,2,3$, are dominated by functions $b_i(x)$, $i=0,1,2,3$, which all have finite variance.
    \item[\textbf{A3}] The KL divergence $D(f(x;\bar\theta)||f(x;\theta))$, where
    \begin{align}
        D(f(x)||g(x))\triangleq \int f(x)\ln\frac{f(x)}{g(x)}\d x,
    \end{align}
    has a unique minimum with respect to $\theta$ at $\theta=\bar{\theta}$.
\item[\textbf{A4}] The expectation $\E\big[\frac{\partial^2}{\partial\theta^2}\ln f(x;\bar\theta)\big]$ is non-zero.
\end{itemize}
The assumption A0 is sufficient (but not necessary) for the convergence of the integral in~\eqref{eqn:MSEHATMLdefinition}. Under the assumptions A1-A3, it can be shown that (See \cite[Theorem 2.1]{White1981}) the ML estimate $\hat\theta\triangleq\arg \max_{\theta} \ln f (\mathbf{x};\theta)$ is consistent, i.e., $\hat\theta\overset{\text{a.s.}}{\rightarrow} \bar{\theta}$ as $N\rightarrow \infty$ . Then, we have
\begin{align}
   \widehat{\MSE}_{\mathrm{ML}}(\bar{\theta})\rightarrow \MSE_{\mathrm{ML}}(\bar{\theta})\label{eqn:ExpressionGoestoTrueMSE}
\end{align}
as $N\rightarrow\infty$, i.e., the finite support version of the MSE expression $\widehat{\MSE}_{\mathrm{ML}}(\bar{\theta})$ in~\eqref{eqn:MSEHATMLdefinition}  converges to the true large sample asymptotic MSE of the ML estimate $\hat{\theta}$ as $N\rightarrow\infty$. If the ML estimate is also asymptotically efficient, then we have
\begin{align}
   \widehat{\MSE}_{\mathrm{ML}}(\bar{\theta})\rightarrow\C(\bar{\theta}),\label{eqn:ExpressionGoestoCRLB}
\end{align}
as $N\rightarrow\infty$ where $\C(\bar{\theta})\triangleq\I^{-1}(\bar{\theta})$ with $\C(\bar{\theta})$ and $\I(\bar\theta)$ denoting the CRLB and the Fisher information matrix, respectively, at the true parameter value $\bar{\theta}$.
\end{proposition}
\textbf{Proof:} The proof is given in \ref{app:ProofforCRLB}.
\hfill$\blacksquare$

\subsection{Relationship to MCRLB}\label{sec:relationtoMCRB}
In this section, we consider the misspecified ML (MML) estimation~\cite{Richmond2015_MCRB,white82} for a scalar parameter $\theta\in\mathbb{R}$. In order to use the asymptotic results for the MML estimate $\hat{\theta}$, we are going to assume that the elements $x_n$, $n=0,\ldots,N-1$, of the measurement vector $\mathbf{x}$ are independent and identically distributed as $x_n\sim \bar{f}(x_n)$ where $\bar{f}(\cdot)$ denotes the true measurement distribution. The true distribution for the measurement vector $\mathbf{x}$ is then given as
\begin{align}
    \bar{f}(\mathbf{x})=\prod_{n=0}^{N-1} \bar{f}(x_n).\label{eqn:iid-truelikelihood}
\end{align}
We assume that MML estimate $\hat{\theta}$ is calculated by maximizing the assumed likelihood $f(\mathbf{x};\theta)$ given in~\eqref{eqn:iid-likelihood}. The relationship of the suggested variance expression $\widehat{V}_{\hat\theta}(\theta)$ to MCRLB~\cite{FortunatiGGR2017,Richmond2015_MCRB} is given in the following proposition.
\begin{proposition}\label{prop:CaseofMCRLB}
\begin{itemize} Assume that
    \item[\textbf{A0}] The parameter $\theta$ has finite support, i.e., $\theta\in[\theta_{\mathrm{min}},\theta_{\mathrm{max}}]$.
    \item[\textbf{A1}] The first three derivatives of $\ln f(x;\theta)$ with respect to $\theta$ exist for all $\theta$ and are continuous with respect to $\theta$.
    \item[\textbf{A2}] For every $\theta$, the functions $|\frac{\partial^i}{\partial\theta^i}\ln f(x;\theta)|$, $i=0,1,2,3$, are dominated by functions $b_i(x)$, $i=0,1,2,3$, which all have finite variance with respect to the true measurement distribution $\bar{f}(x)$.
    \item[\textbf{A3}] The KL divergence $D(\bar{f}(x)||f(x;\theta))$ has a unique minimum with respect to $\theta$ at $\theta=\theta_*\in(\theta_{\mathrm{min}},\theta_{\mathrm{max}})$.\footnote{Note that the existence of the KL divergence $D(\bar{f}(x)||f(x;\theta))$ necessitates additionally the existence of the $E_{\bar{f}}[\ln\bar{f}(x)]$, which we implicitly assume for the sake of conceptual simplicity. We may eliminate the need for the existence of $E_{\bar{f}}[\ln\bar{f}(x)]$ by stating this assumption differently as in~\cite{White1981,Vuong1986,FortunatiGG2017}.}
    \item[\textbf{A4}] The expectation $\E_{\bar{f}}\big[\frac{\partial^2}{\partial\theta^2}\ln f(x;\theta_*)\big]$ is non-zero.
\end{itemize}
The assumption A0 is sufficient (but not necessary) for the convergence of the integral in~\eqref{eqn:VHATdefinition}. Under the assumptions A1-A3 it can be shown that (See \cite[Theorem 2.1]{White1981}) the MML estimate $\hat\theta\triangleq\arg \max_{\theta} \ln f(\mathbf{x};\theta)$ is misspecified consistent, i.e., $\hat\theta\overset{\text{a.s.}}{\rightarrow} \theta_*$ as $N\rightarrow \infty$ . Then, we have
\begin{align}
   \widehat{V}_{\mathrm{MML}}(\theta_*)\rightarrow V_{\mathrm{MML}}(\theta_*)\label{eqn:ExpressionGoestoTrueV}
\end{align}
as $N\rightarrow\infty$, i.e., the finite support version of the expression $\widehat{V}_{\mathrm{MML}}(\theta_*)$ converges to the true large sample asymptotic variance $V_{\mathrm{MML}}(\theta_*)$ of the MML estimate $\hat{\theta}$ as $N\rightarrow\infty$. If the MML estimate is also asymptotically misspecified efficient, then we have
\begin{align}
   \widehat{V}_{\mathrm{MML}}(\theta_*)\rightarrow\frac{\mathcal{B}(\theta_*)}{\mathcal{A}^2(\theta_*)}\label{eqn:ExpressionGoestoMCRLB}
\end{align}
as $N\rightarrow\infty$ where the quantity $\frac{\mathcal{B}(\theta_*)}{\mathcal{A}^2(\theta_*)}$ is the MCRLB and
    \begin{equation}
        \mathcal{A}(\theta_*)\triangleq\,\E_{\bar{f}}\bigg[\frac{\partial^2 }{\partial\theta^2}\ln f(x;\theta_*)\bigg], \quad
        \mathcal{B}(\theta_*)\triangleq\,\E_{\bar{f}}\bigg[\bigg(\frac{\partial }{\partial \theta}\ln f(x;\theta_*)\bigg)^2\bigg].
    \end{equation}
\end{proposition}
\textbf{Proof:} The proof is given in \ref{app:ProofforMCRLB}.
\hfill$\blacksquare$

Note that according to the proposition, $\widehat{V}(\theta_*)$ converges to the asymptotic variance $V_{\mathrm{MML}}(\theta_*)$ of the MML estimate $\hat{\theta}$. When the true measurement distribution $\bar{f}(\cdot)$ admits the same parameterization as the assumed measurement distribution $f(\cdot;\theta)$ with the true parameter value $\theta=\bar\theta$, i.e., $\bar{f}(x)=\bar{f}(x;\bar\theta)$, then we might predict the MSE performance of the MML estimator $\hat{\theta}$ as
\begin{align}
    \widehat{\MSE}_{\mathrm{MML}}(\bar\theta)\triangleq \widehat{V}_{\mathrm{MML}}(\theta_*)+(\theta_*-\bar\theta)^2,
\end{align}
which would converge to the true MSE of the MML estimator as $N\rightarrow\infty$ if the assumptions of Proposition~\ref{prop:CaseofMCRLB} are satisfied.

\subsection{Relationship to ZZB}~\label{sec:relationtoZZB}
We consider a Bayesian estimation problem where the parameter $\theta$ is assigned with the prior distribution $f(\theta)$. The MAP estimate of $\theta$ can then be defined as follows.
\begin{align}
    \hat{\theta}\triangleq \arg\max_{\theta} f(\mathbf{x}|\theta)f(\theta)\label{eqn:MAPestimatorDefinition}
\end{align}
where the likelihood $f(\mathbf{x};\theta)$ is shown with the conditioning notation as $f(\mathbf{x}|\theta)$ since $\theta$ is now a random variable.
Note that the MAP estimator given above corresponds to an IDE with the objective function $\mathcal{L}(\mathbf{x};\theta)\triangleq f(\mathbf{x}|\theta)f(\theta)$. The true MSE of the MAP estimator is given as
\begin{subequations}
\begin{align}
\MSE_{\mathrm{MAP}}\triangleq& \int \int (\hat{\theta}-\theta)^2 f(\mathbf{x}|\theta)\d \mathbf{x} f(\theta)\d \theta\\
=&\,\E\big[\E\big[(\hat\theta-\theta)^2\big|\theta\big]\big]\label{eqn:averageMSEdoubleexpectation}\\
=&\,\E\big[\MSE_{\mathrm{MAP}}(\theta)\big],\label{eqn:averageMSEdefinition}
\end{align}
\end{subequations}
where the outer expectation in~\eqref{eqn:averageMSEdoubleexpectation} is with respect to the random variable $\theta$ and $\MSE_{\mathrm{MAP}}(\theta)$ denotes the true MSE of the MAP estimator when $\theta$ is given, i.e.,
\begin{align}
    \MSE_{\mathrm{MAP}}(\theta)\triangleq\E\big[(\hat\theta-\theta)^2\big|\theta\big],
\end{align}
where the expectation is only with respect to the noisy measurements $\mathbf{x}$ given $\theta$. Since the problem becomes a non-random parameter estimation problem when $\theta$ is given, we can predict $\MSE_{\mathrm{MAP}}(\theta)$ of the MAP estimator using~\eqref{eqn:MSEHATdefinition} as follows.
\begin{align}
    \widehat{\MSE}_{\mathrm{MAP}}(\theta)=2\int_{-\infty}^\infty
\! |\epsilon| P\Big( f(\mathbf{x}|\theta+2\epsilon)f(\theta+2\epsilon)\ge f(\mathbf{x}|\theta)f(\theta)\Big|\theta\Big)\d\epsilon.\label{eqn:MSEMAPtheta}
\end{align}
By substituting the MSE estimate $\widehat{\MSE}_{\mathrm{MAP}}(\theta)$ in~\eqref{eqn:MSEMAPtheta} into the place of $\MSE_{\mathrm{MAP}}(\theta)$ in~\eqref{eqn:averageMSEdefinition} we can predict the overall MSE of the MAP estimate as follows.
\begin{subequations}
\begin{align}
&\hspace{-0.5cm}\widehat{\MSE}_{\mathrm{MAP}}\triangleq \E[\widehat{\MSE}_{\mathrm{MAP}}(\theta)]\\
\triangleq&\,2\int_{-\infty}^\infty f(\theta)\int_{-\infty}^\infty|\epsilon|P\Big( f(\mathbf{x}|\theta+2\epsilon)f(\theta+2\epsilon)\ge f(\mathbf{x}|\theta)f(\theta)\Big|\theta\Big)\d\epsilon\d\theta\\
=&\int_{-\infty}^\infty \int_{-\infty}^\infty|\epsilon|f(\theta)P\Big( f(\mathbf{x}|\theta+2\epsilon)f(\theta+2\epsilon)\ge f(\mathbf{x}|\theta)f(\theta)\Big|\theta\Big)\d\epsilon\d\theta \nonumber \\
&+\int_{-\infty}^\infty \int_{-\infty}^\infty|\epsilon|f(\theta-2\epsilon)P\Big( f(\mathbf{x}|\theta-2\epsilon)f(\theta-2\epsilon)\ge f(\mathbf{x}|\theta)f(\theta)\Big|\theta\Big)\d\epsilon\d\theta\\
=&\int_{-\infty}^\infty \int_{-\infty}^\infty|\epsilon|f(\theta)P\Big( f(\mathbf{x}|\theta+2\epsilon)f(\theta+2\epsilon)\ge f(\mathbf{x}|\theta)f(\theta)\Big|\theta\Big)\d\epsilon\d\theta \nonumber \\
&+\int_{-\infty}^\infty \int_{-\infty}^\infty|\epsilon|f(\theta+2
\epsilon)P\Big( f(\mathbf{x}|\theta)f(\theta)\ge f(\mathbf{x}|\theta+2\epsilon)f(\theta+2\epsilon)\Big|\theta+2\epsilon\Big)\d\epsilon\d\theta\\
=&\int_{-\infty}^\infty \int_{-\infty}^\infty|\epsilon|(f(\theta)+f(\theta+2\epsilon))\nonumber\\
&\times\Big[
\pi_1 P\Big( \pi_2f(\mathbf{x}|\theta+2\epsilon)\ge \pi_1 f(\mathbf{x}|\theta)\Big|\theta\Big)+\pi_2P\Big( \pi_1 f(\mathbf{x}|\theta)\ge \pi_2 f(\mathbf{x}|\theta+2\epsilon) \Big|\theta+2\epsilon\Big)\Big]\d\epsilon\d\theta\\
=&\int_{-\infty}^\infty \int_{-\infty}^\infty|\epsilon|(f(\theta)+f(\theta+2\epsilon))P_{\min}^e(\theta,\theta+2\epsilon)\d\epsilon\d\theta \\
=&\,2\int_{0}^\infty \int_{-\infty}^\infty\epsilon(f(\theta)+f(\theta+2\epsilon)) P_{\min}^e(\theta,\theta+2\epsilon)\d\theta\d\epsilon \nonumber \\
=&\frac{1}{2}\int_{0}^\infty \int_{-\infty}^\infty\epsilon(f(\theta)+f(\theta+\epsilon)) P_{\min}^e(\theta,\theta+\epsilon)\d\theta\d\epsilon,\label{eqn:averageMSEZZB}
\end{align}
\end{subequations}
where $P_{\min}^{e}(\theta_1,\theta_2)$ is the minimum probability of error for the binary hypothesis testing problem given below.
\begin{subequations}
\begin{align}
\H_1 :&\,\, \mathbf{x}\sim f(\mathbf{x}|\theta_1),\\
\H_2 :&\,\, \mathbf{x}\sim f(\mathbf{x}|\theta_2),
\end{align}
\end{subequations}
with the prior hypothesis probabilities $P(\H_1)=\pi_1$ and $P(\H_2)=\pi_2=1-\pi_1$ where
\begin{align}
\pi_1\triangleq& \frac{f(\theta_1)}{f(\theta_1)+f(\theta_2)},&\pi_2\triangleq& \frac{f(\theta_2)}{f(\theta_1)+f(\theta_2)}.
\end{align}
The expression~\eqref{eqn:averageMSEZZB} can be seen to be the ZZB (See~\cite[Eqn.\  (14)]{BellSEVT:1997}) without the so-called valley filling function. As a result $\widehat{\MSE}_{\mathrm{MAP}}$ calculated using~\eqref{eqn:MSEHATdefinition} in a Bayesian framework is equal to the ZZB. Note that this equality is satisfied irrespective of whether the objective function $f(\mathbf{x}|\theta)f(\theta)$ satisfies the assumptions of Theorem~\ref{thm:IDEV} or not.  If the objective function $f(\mathbf{x}|\theta)f(\theta)$, which is actually the joint density $f(\mathbf{x},\theta)$ of $\mathbf{x}$ and $\theta$, also satisfies the conditions of Theorem~\ref{thm:IDEV}, then this would mean that $\MSE_{\mathrm{MAP}}(\theta)=\widehat{\MSE}_{\mathrm{MAP}}(\theta)$ for all $\theta\in\mathbb{R}$ and hence $\MSE_{\mathrm{MAP}}=\widehat{\MSE}_{\mathrm{MAP}}=\mathrm{ZZB}$ and hence ZZB would have to be tight, i.e., ZZB would have to be equal to the true average MSE of the MAP estimate $\hat{\theta}$.  As a result, the conditions of Theorem~\ref{thm:IDEV} are also a set of sufficient conditions for ZZB to be tight.

\section{Extension to the Case with Nuisance Parameters}\label{sec:multipleparameters}
Suppose now that we have $J>1$ unknown scalar parameters, i.e., $\boldsymbol{\theta}\in\mathbb{R}^J$, and we would like to estimate only one of them while keeping the others as unknown nuisance parameters. Without loss of generality we assume that we would like to estimate $\theta_1$ while treating the other parameters $\theta_2,\ldots,\theta_J$ as nuisance parameters. We can express the estimate $\hat{\theta}_1$ for  $\theta_1$ as
    \begin{align}
        \hat\theta_1\triangleq[\boldsymbol{\hat\theta}]_1=& \arg\max_{\theta_1}\Big[\underbrace{\max_{\theta_{\backslash 1}}\L(\mathbf{x};\boldsymbol{\theta})}_{\triangleq \L_1(\mathbf{x},\theta_1)}\Big] = \arg\max_{\theta_1} \L_1(\mathbf{x},\theta_1),\label{eqn:multipleparameters}
    \end{align}
where $\boldsymbol{\theta}_{\backslash 1}\triangleq \left[\theta_2\,\,\theta_3\,\cdots\,\theta_J\right]^\t$. If we assume that the function $\L_1(\mathbf{x},\theta_1)$ defined as $\L_1(\mathbf{x},\theta_1)\triangleq\max_{\boldsymbol{\theta}_{\backslash 1}}\L(\mathbf{x};\boldsymbol{\theta})$ satisfies the conditions in Theorem~\ref{thm:IDEV}, applying the result of Remark~\ref{cor:MSE} to the IDE in~\eqref{eqn:multipleparameters} would give
\begin{align}
    \widehat{\MSE}(\bar{\theta}_1)=&\,2\int_{-\infty}^\infty|\epsilon|P\left(\L_1(\mathbf{x},\bar\theta_1+2\epsilon)\ge\L_1(\mathbf{x},\bar\theta_1)\right)\d\epsilon,\nonumber\\
    =&\,2\int_{-\infty}^\infty|\epsilon| P\Big(\max_{\boldsymbol{\theta}_{\backslash 1}}\L(\mathbf{x};\bar\theta_1+2\epsilon,\boldsymbol{\theta}_{\backslash 1})\ge\max_{\boldsymbol{\theta}_{\backslash 1}}\L(\mathbf{x};\bar\theta_1,\boldsymbol{\theta}_{\backslash 1})\Big)\d\epsilon.\label{eqn:actualmultipleparametercase}
\end{align}
With the selection $\L(\mathbf{x};\boldsymbol\theta)\triangleq f(\mathbf{x};\boldsymbol\theta)\ge 0$, we can obtain the MSE of the ML estimate
$\hat\theta_1$ of $\theta_1$ similarly to Corollary~\ref{cor:MLcase} from~\eqref{eqn:actualmultipleparametercase} as
\begin{align}
    \widehat{\MSE}_{\mathrm{ML}}(\bar{\theta}_1)=&\,2\int_{-\infty}^\infty|\epsilon| P\bigg(\myfrac[2pt]{\max_{\boldsymbol{\theta}_{\backslash 1}}f(\mathbf{x};\bar\theta_1+2\epsilon,\boldsymbol{\theta}_{\backslash 1})}{\max_{\boldsymbol{\theta}_{\backslash 1}}f(\mathbf{x};\bar\theta_1,\boldsymbol{\theta}_{\backslash 1})}\ge 1\bigg)\d\epsilon,\label{eqn:MLmultipleparametercase}
\end{align}
connecting the MSE of the ML estimator to the error probability of a generalized likelihood ratio test (GLRT) (instead of a likelihood ratio test) in the presence of nuisance parameters \cite{Kay_DetectionTheory}.

In Section~\ref{sec:ParametricMeanNuisance} below, we are going to investigate the expression~\eqref{eqn:MLmultipleparametercase} further on the specific case of the parametric mean model and make approximations to facilitate its calculation, which are later extended to the general case in a remark.

\section{Application to ML Estimation with the Parametric Mean Model with Gaussian Noise}\label{sec:parameterizedmean}
We consider ML estimator with the measurement model given as
\begin{align}
    \mathbf{x}=\mathbf{m}(\boldsymbol{\bar\theta})+\mathbf{v}, \label{eq:parametricMeanModel}
\end{align}
where $\mathbf{v}\sim\CN(\mathbf{v};\mathbf{0},\sigma^2\mathbf{I}_N)$ represents the measurement noise and the manifold function $\mathbf{m}:\mathbb{R}^J\rightarrow\mathbb{C}^{N}$ is, in  general, a complex-valued function of the unknown parameter vector $\boldsymbol{\theta}\in\mathbb{R}^J$. The measurement model in \eqref{eq:parametricMeanModel} is widely used in signal processing applications. For example, a linear manifold function $\mathbf{m}(\boldsymbol{\bar\theta}) = \mathbf{H}\boldsymbol{\bar\theta}$ may represent a multi-input multi-output (MIMO) communication system; a non-linear  manifold function may represent the array response in the direction of arrival estimation problems \cite{Kay_EstimationTheory}.

The likelihood function for an arbitrary $\boldsymbol{\theta}$ is given as
\begin{align}
    f(\mathbf{x};\boldsymbol{\theta})=\CN(\mathbf{x};\mathbf{m}(\boldsymbol{\theta}),\sigma^2\mathbf{I}_N).\label{eqn:ParameterizedMeanModelLikelihood}
\end{align}
We investigate the cases of a scalar parameter with and without nuisance parameters in different subsections below. In order to calculate the predicted MSE values we will need the following log-likelihood ratio expression.
\begin{align}
\ln\myfrac[1.5pt]{f(\mathbf{x};\boldsymbol{\theta})}{f(\mathbf{x};\boldsymbol{\bar\theta})}=&\frac{1}{\sigma^2}\big(2\Re\{\mathbf{\tilde{m}}^\h (\boldsymbol{\theta};\boldsymbol{\bar\theta}) (\mathbf{x}-\mathbf{m}(\bar\theta))\}-\|\mathbf{\tilde{m}}(\boldsymbol{\theta};\boldsymbol{\bar\theta})\|^2\big),
\end{align}
where
$
\mathbf{\tilde{m}}(\boldsymbol{\theta}_1;\boldsymbol{\theta}_2)\triangleq \mathbf{m}(\boldsymbol{\theta}_1)-\mathbf{m}(\boldsymbol{\theta}_2).\label{eqn:mtildeDefinition}
$

\subsection{Case of a Scalar Parameter with No Nuisance Parameters}\label{sec:parametricmeansingle}
Suppose now that we have a scalar parameter $\theta$ with the true value $\bar\theta$ ($J=1$). Note that this case can also be interpreted to be the case when we have multiple parameters $\boldsymbol{\theta}=[\theta_1\,\,\,\boldsymbol{\theta}_{\backslash 1}]$ and the true values $\boldsymbol{\bar\theta}_{\backslash 1}$ of the nuisance parameters $\boldsymbol{\theta}_{\backslash 1}$ are perfectly known.
We can evaluate the probability in~\eqref{eqn:MSEHATMLdefinition} as
\begin{subequations}
    \begin{align}
        P\left(\myfrac[1.5pt]{f(\mathbf{x};\bar{\theta}+2\epsilon)}{f(\mathbf{x};\bar{\theta})}\ge 1\right)&=P\left( \ln\myfrac[1.5pt]{f(\mathbf{x};\bar{\theta}+2\epsilon)}{f(\mathbf{x};\bar{\theta})}\ge 0\right)=P\left(2\Re\{\mathbf{\tilde{m}}^\h(\bar{\theta}+2\epsilon;\bar\theta)\mathbf{v}\} \ge \|\mathbf{\tilde{m}}(\bar{\theta}+2\epsilon;\bar\theta)\|^2\right)\\
        &=\N_{\mathrm{\mathrm{ccdf}}}\left(\|\mathbf{\tilde{m}}(\bar{\theta}+2\epsilon;\bar\theta)\|^2;0,2\sigma^2\|\mathbf{\tilde{m}}(\bar{\theta}+2\epsilon;\bar\theta)\|^2\right)\label{eqn:singleparameterprobability}\\
        &=\N_{\mathrm{\mathrm{ccdf}}}\left(\|\mathbf{\tilde{m}}(\bar{\theta}+2\epsilon;\bar\theta)\|;0,2\sigma^2\right),\label{eqn:probabilityforscalarcase}
    \end{align}
\end{subequations}
under the assumption that $\|\mathbf{\tilde{m}}(\bar{\theta}+2\epsilon;\bar\theta)\|\neq 0$, where $\N_{\mathrm{\mathrm{ccdf}}}(\mathbf{x};\boldsymbol{\mu},\boldsymbol{\Sigma})$ denotes the complementary cumulative distribution function ($\mathrm{\mathrm{ccdf}}$) of a real Gaussian random vector with mean $\boldsymbol{\mu}$ and covariance $\boldsymbol{\Sigma}$ evaluated at $\mathbf{x}$. Assuming that $\|\mathbf{\tilde{m}}(\bar{\theta}+2\epsilon;\bar\theta)\|\neq 0$ for almost all $\epsilon\in\mathbb{R}$, we can substitute this probability expression into~\eqref{eqn:MSEHATMLdefinition} to get
\begin{align}
\widehat{\MSE}_{\mathrm{ML}}(\bar{\theta})=\,2\int_{-\infty}^\infty|\epsilon|\N_{\mathrm{ccdf}}\left(\|\mathbf{\tilde{m}}(\bar{\theta}+2\epsilon,\bar\theta)\|;0,2\sigma^2\right)\d\epsilon.\label{eqn:scalarperformancepredictor}
\end{align}
\begin{remark}
If both the function $\mathbf{m}(\cdot)$ and the measurement noise $\mathbf{v}$ are real-valued, i.e., if we have $\mathbf{m}:\mathbb{R}\rightarrow \mathbb{R}^N$ and $\mathbf{v}\sim \N(\mathbf{v};\mathbf{0},\sigma^2\mathbf{I}_N)$, then, instead of~\eqref{eqn:scalarperformancepredictor}, one needs to use
\begin{align}
\widehat{\MSE}_{\mathrm{ML}}(\bar{\theta})=\,2\int_{-\infty}^\infty|\epsilon|\N_{\mathrm{ccdf}}\left(\|\mathbf{\tilde{m}}(\bar{\theta}+2\epsilon,\bar\theta)\|;0,4\sigma^2\right)\d\epsilon.~\label{eqn:scalarperformancepredictorwithrealnoise}
\end{align}
Using this expression amounts to replacing the variance $\sigma^2$ in~\eqref{eqn:scalarperformancepredictor} with $2\sigma^2$.\hfill$\blacksquare$
\end{remark}

Note that the likelihood~\eqref{eqn:ParameterizedMeanModelLikelihood} does not satisfy the conditions of Theorem~\ref{thm:IDEV} and its corollary in general except for some trivial cases, e.g., the case of linear or affine manifold function $\mathbf{m}(\theta)$. As a result, the predicted MSE expressions in~\eqref{eqn:scalarperformancepredictor} and~\eqref{eqn:scalarperformancepredictorwithrealnoise} are expected to be only an approximate estimate of the true MSE of the ML estimator. Furthermore, a closed form solution rarely exists for the integrals in~\eqref{eqn:scalarperformancepredictor} and \eqref{eqn:scalarperformancepredictorwithrealnoise}. Therefore, numerical integration methods have to be used as shown in Example~\ref{ex:frequencyEstimation} below.

The relations in~\eqref{eqn:scalarperformancepredictor} and~\eqref{eqn:scalarperformancepredictorwithrealnoise} provide some insight on the suggested MSE expression. As $\|\mathbf{\tilde{m}}(\bar{\theta}+2\epsilon,\bar\theta)\|$, which is the norm of the  difference between manifold vectors $\mathbf{m}(\bar\theta+2\epsilon)$ and $\mathbf{m}(\bar\theta)$, gets larger, it should be easier to accurately estimate $\theta$ and we get a smaller predicted MSE value (since the function $\N_{\mathrm{ccdf}}(\cdot)$ monotonically decreases as its argument gets larger). Also, it is interesting to see that, for the simplest case $\mathbf{{m}(\theta)} \triangleq \theta$, the expression for the predicted MSE in  \eqref{eqn:scalarperformancepredictor} simplifies to;
\begin{align}
\widehat{\MSE}_{\mathrm{ML}}(\bar{\theta})&=\,2\int_{-\infty}^\infty|\epsilon|\N_{\mathrm{ccdf}}\left(|2\epsilon|;0,2\sigma^2\right)\d\epsilon =\frac{\sigma^2}{2},
\end{align}
which can be obtained by  integration by parts. This result is expected, as the corresponding ML estimator is $\hat{\theta}=\Re\{x\}$, hence, the true MSE must be equal to half of the noise variance. We have an exact result since the objective function is a Gaussian likelihood satisfying the conditions of Theorem~\ref{thm:IDEV}. We finally consider the following example in order to illustrate the practical simplicity of the expression~\eqref{eqn:scalarperformancepredictor}.
\begin{example}[Frequency estimation using ML]\label{ex:frequencyEstimation} Consider the following signal model.
\begin{align}
    x_n=Ae^{j\bar{\omega} n}+v_n,\quad n=0,\ldots,N-1,
\end{align}
where $A\in\mathbb{C}$ is the known complex amplitude;  $\bar\omega\in[-\pi,\pi]$ is the unknown true frequency to be estimated using the ML estimator; $v_n\sim\CN(v_n;0,\sigma^2)$, $n=0,\ldots,N-1$, is the white measurement noise. MSE of the ML estimator based on the measurements $x_n$, $n=0,\ldots,N-1$, can be calculated with the Matlab function given in Figure~\ref{fig:MatlabCode}, which involves only three lines of code. A sample run can be made using the command \mcode{MSE_ML_frequency(pi/2,1,1,16)} for the true frequency value $\bar\omega=\pi/2$, amplitude  $A=1$, noise variance $\sigma^2=1$ and number of samples $N=16$ gives $\widehat\MSE_{\mathrm{ML}}(\frac{\pi}{2})=$\mcode{6.417e-4}\,$\mathrm{rad}^2$. Note that the function \mcode{calculateMSEhat(.)} in Figure~\ref{fig:MatlabCode} can be used for predicting the MSE performance of the ML estimator for any measurement model of type~\eqref{eq:parametricMeanModel} for a scalar parameter $\theta\in[\theta_{\mathrm{min}},\theta_{\mathrm{max}}]$.
\end{example}
\begin{figure}[tb]
\vspace{-1cm}\begin{lstlisting}[mathescape]
function MSEhatML = MSE_ML_frequency(tw,A,sigma2,N)
% Calculate MSEhatML for ML frequency estimate
%     tw: true value of the frequency (1x1) (rad/sec) (-pi < tw < pi)
%      A: known complex amplitude (1x1)
% sigma2: known complex normal noise variance (1x1)
%      N: number of samples (1x1)
m = @(w) A*exp (1i*(0:N-1)'*w);
MSEhatML = calculateMSEhat(m,tw,-pi,pi,sigma2);

function MSEhatML = calculateMSEhat(m,tt,tmin,tmax,sigma2)
% Calculates the MSEhat for the parameterized mean model
%      m: function handle: m(.) takes a 1xL array of parameter values [theta1 theta2 ... thetaL]
%       : and it returns the Nm x L matrix [m(theta1) m(theta2) ... m(thetaL)]
%     tt: true theta value (1x1)
%   tmin: minimum value of theta (1x1)
%   tmax: maximum value of theta (1x1)
% sigma2: complex normal noise variance (1x1)
MSEhatML = 2*integral(@(e)abs(e).*(1-normcdf(vecnorm(m(tt+2*e)-m(tt),2,1)/sqrt(2*sigma2))),...
         (tmin-tt)/2,(tmax-tt)/2);
\end{lstlisting}
\caption{A Matlab code (R2021b) for predicting the MSE of the ML estimator for the frequency estimation problem.}
\label{fig:MatlabCode}
\end{figure}

\subsection{Case of a Scalar Parameter with Nuisance Parameters}\label{sec:ParametricMeanNuisance}
When some nuisance parameters exist, we consider the case in Section~\ref{sec:multipleparameters} and use the MSE expression in~\eqref{eqn:MLmultipleparametercase}. Unfortunately it is analytically difficult  to calculate the maxima and the probabilities in the integrands on the right hand side of~\eqref{eqn:MLmultipleparametercase} exactly. In the following, we are going to make some approximations to facilitate the calculation. Similar approximations can also be made for the more general case in~\eqref{eqn:actualmultipleparametercase} (See Remark~\ref{rem:approximationsforgeneralcase} below).
    \begin{align}
        \myfrac[1.5pt]{\max_{\boldsymbol{\theta}_{\backslash 1}}f(\mathbf{x};\bar\theta_1+2\epsilon,\boldsymbol{\theta}_{\backslash 1})}{\max_{\boldsymbol{\theta}_{\backslash 1}}f(\mathbf{x};\bar\theta_1,\boldsymbol{\theta}_{\backslash 1})}\approx&\, \myfrac[1.5pt]{\max_{\boldsymbol{\theta}_{\backslash 1}}f(\mathbf{x};\bar\theta_1+2\epsilon,\boldsymbol{\theta}_{\backslash 1})}{f(\mathbf{x};\bar\theta_1,\boldsymbol{\bar\theta}_{\backslash 1})}=\max_{\boldsymbol{\theta}_{\backslash 1}}\myfrac[1.5pt]{f(\mathbf{x};\bar\theta_1+2\epsilon,\boldsymbol{\theta}_{\backslash 1})}{f(\mathbf{x};\bar\theta_1,\boldsymbol{\bar\theta}_{\backslash 1})}\approx \max_{\boldsymbol{\theta}_{\backslash 1}\in\boldsymbol{\Theta}_{\backslash 1}}\myfrac[1.5pt]{f(\mathbf{x};\bar\theta_1+2\epsilon,\boldsymbol{\theta}_{\backslash 1})}{f(\mathbf{x};\bar\theta_1,\boldsymbol{\bar\theta}_{\backslash 1})},\label{eqn:SetofGridValues}
    \end{align}
 where the first approximation in~\eqref{eqn:SetofGridValues} is made by assuming that the maximum in the denominator of the left hand side is achieved approximately at the true values of the nuisance parameters, i.e., at $\boldsymbol{\theta}_{\backslash 1}=\boldsymbol{\bar\theta}_{\backslash 1}$, which is reasonable under asymptotic conditions.
The set
$\boldsymbol{\Theta}_{\backslash 1}\triangleq\{\boldsymbol{\theta}_{\backslash 1}^1,\boldsymbol{\theta}_{\backslash 1}^2,\ldots,\boldsymbol{\theta}_{\backslash 1}^{N_\theta} \}$ appearing in~\eqref{eqn:SetofGridValues} is a set of grid points including the true value $\boldsymbol{\bar\theta}_{\backslash 1}$ of $\boldsymbol{\theta}_{\backslash 1}$. Using these approximations, we can approximate the probability in~\eqref{eqn:MLmultipleparametercase} as  \begin{subequations}
    \begin{align}
        P\bigg(&\myfrac[1.5pt]{\max_{\boldsymbol{\theta}_{\backslash 1}}f(\mathbf{x};\bar\theta_1+2\epsilon,\boldsymbol{\theta}_{\backslash 1})}{\max_{\boldsymbol{\theta}_{\backslash 1}}f(\mathbf{x};\bar\theta_1,\boldsymbol{\theta}_{\backslash 1})}\ge 1\bigg) \approx\, P\bigg(\max_{\boldsymbol{\theta}_{\backslash 1}\in\boldsymbol{\Theta}_{\backslash 1}}\myfrac[1.5pt]{f(\mathbf{x};\bar\theta_1+2\epsilon,\boldsymbol{\theta}_{\backslash 1})}{f(\mathbf{x};\bar\theta_1,\boldsymbol{\bar\theta}_{\backslash 1})}\ge 1\bigg),\\
        =&\, P\bigg(\max_{\boldsymbol{\theta}_{\backslash 1}\in\boldsymbol{\Theta}_{\backslash 1}}\ln\myfrac[1.5pt]{f(\mathbf{x};\bar\theta_1+2\epsilon,\boldsymbol{\theta}_{\backslash 1})}{f(\mathbf{x};\bar\theta_1,\boldsymbol{\bar\theta}_{\backslash 1})}\ge 0\bigg)\\
        =&\, P\bigg(\max_{\boldsymbol{\theta}_{\backslash 1}\in\boldsymbol{\Theta}_{\backslash 1}}
        \Big[2\Re\{\mathbf{\tilde{m}}^\h (\bar\theta_1+2\epsilon,\boldsymbol{\theta}_{\backslash 1};\bar\theta_1,\boldsymbol{\bar\theta}_{\backslash 1}) \mathbf{v}\}-\|\mathbf{\tilde{m}}(\bar\theta_1+2\epsilon,\boldsymbol{\theta}_{\backslash 1};\bar\theta_1,\boldsymbol{\bar\theta}_{\backslash 1})\|^2\Big]
        \ge 0\bigg)\label{eqn:multipleparameterderivation1}\\
        =&\,1-P\bigg(\max_{\boldsymbol{\theta}_{\backslash 1}\in\boldsymbol{\Theta}_{\backslash 1}}
        \Big[2\Re\{\mathbf{\tilde{m}}^\h (\bar\theta_1+2\epsilon,\boldsymbol{\theta}_{\backslash 1};\bar\theta_1,\boldsymbol{\bar\theta}_{\backslash 1}) \mathbf{v}\}-\|\mathbf{\tilde{m}}(\bar\theta_1+2\epsilon,\boldsymbol{\theta}_{\backslash 1};\bar\theta_1,\boldsymbol{\bar\theta}_{\backslash 1})\|^2\Big]
        \le 0\bigg)\label{eqn:multipleparameterderivation2}
    \end{align}
\end{subequations}
where
$
    \mathbf{\tilde{m}}(\theta_1^1,\boldsymbol{\theta}_{\backslash 1}^1;\theta_1^2,\boldsymbol{\theta}_{\backslash 1}^2)\triangleq \mathbf{m}(\theta_1^1,\boldsymbol{\theta}_{\backslash 1}^1)-\mathbf{m}(\theta_1^2,\boldsymbol{\theta}_{\backslash 1}^2).
$

Let us now define the  matrix $\mathbf{\widetilde{M}}_{\epsilon}\in\mathbb{R}^{N\times N_\theta}$ and the vector $\boldsymbol{\tilde\mu}_{\epsilon}\in\mathbb{R}^{N_\theta}$ as

\begin{equation}
    \mathbf{\widetilde{M}}_{\epsilon}\triangleq\left[\begin{array}{c}
    \mathbf{\tilde{m}}^\h(\bar\theta_1+2\epsilon,\boldsymbol{\theta}_{\backslash 1}^1;\bar\theta_1,\boldsymbol{\bar\theta}_{\backslash 1})\\
    \vdots\\
    \mathbf{\tilde{m}}^\h(\bar\theta_1+2\epsilon,\boldsymbol{\theta}_{\backslash 1}^{N_\theta};\bar\theta_1,\boldsymbol{\bar\theta}_{\backslash 1})
    \end{array}\right]^\h, \qquad
    \boldsymbol{\tilde\mu}_{\epsilon}\triangleq\left[\begin{array}{c}\big\|
    \mathbf{\tilde{m}}(\bar\theta_1+2\epsilon,\boldsymbol{\theta}_{\backslash 1}^1;\bar\theta_1,\boldsymbol{\bar\theta}_{\backslash 1})\big\|^2\\
    \vdots\\
    \big\|
    \mathbf{\tilde{m}}(\bar\theta_1+2\epsilon,\boldsymbol{\theta}_{\backslash 1}^{N_\theta};\bar\theta_1,\boldsymbol{\bar\theta}_{\backslash 1})\big\|^2
    \end{array}\right].
\end{equation}
We can now write~\eqref{eqn:multipleparameterderivation2} as
\begin{equation}
    P\bigg(\myfrac[1.5pt]{\max_{\boldsymbol{\theta}_{\backslash 1}}f(\mathbf{x};\bar\theta_1+2\epsilon,\boldsymbol{\theta}_{\backslash 1})}{\max_{\boldsymbol{\theta}_{\backslash 1}}f(\mathbf{x};\bar\theta_1,\boldsymbol{\theta}_{\backslash 1})}\ge 1\bigg)
    \approx\,1-P\Big(
    2\Re\{\mathbf{\widetilde{M}}_{\epsilon}^\h\mathbf{v}\}
    \le \boldsymbol{\tilde\mu}_{\epsilon}\Big|\boldsymbol{\bar\theta}\Big)
    =\,\N_{\mathrm{ccdf}}\big(\boldsymbol{\tilde\mu}_{\epsilon};\mathbf{0},
    2\sigma^2\Re\big\{\mathbf{\widetilde{M}}_{\epsilon}^\h\mathbf{\widetilde{M}}_{\epsilon}\big\}\big),\label{eqn:multipleparameterprobability}
\end{equation}
under the assumption that $\|\mathbf{\tilde{m}}(\bar\theta_1+2\epsilon,\boldsymbol{\theta}_{\backslash 1}^i;\bar\theta_1,\boldsymbol{\bar\theta}_{\backslash 1})\|\neq 0$ for $i=1,\ldots,N_\theta$. Note that the inequalities between vector quantities above should be interpreted in an elementwise manner. The probability given in~\eqref{eqn:multipleparameterprobability} is the generalization of the single parameter probability in~\eqref{eqn:singleparameterprobability} to the case of (presence of) nuisance parameters. In fact, when we select the set $\boldsymbol{\Theta}_{\backslash 1}$ as $\boldsymbol{\Theta}_{\backslash 1}=\{\boldsymbol{\bar\theta}_{\backslash 1}\}$, i.e., when we have a grid composed of only the true nuisance parameter $\boldsymbol{\bar\theta}_{\backslash 1}$, the probability~\eqref{eqn:multipleparameterprobability} reduces to the probability~\eqref{eqn:singleparameterprobability}. Moreover since the set $\boldsymbol{\Theta}_{\backslash 1}$ contains the true value $\boldsymbol{\bar\theta}_{\backslash 1}$, the probability~\eqref{eqn:multipleparameterprobability} is always larger than or equal to the probability~\eqref{eqn:singleparameterprobability}. Substituting the result~\eqref{eqn:multipleparameterprobability} into~\eqref{eqn:MLmultipleparametercase} we get
\begin{align}
\widehat{\MSE}(\bar{\theta}_1)=\,2\int_{-\infty}^\infty|\epsilon|\N_{\mathrm{ccdf}}\big(\boldsymbol{\tilde\mu}_{\epsilon};\mathbf{0},
2\sigma^2\Re\big\{\mathbf{\widetilde{M}}_{\epsilon}^\h\mathbf{\widetilde{M}}_{\epsilon}\big\}\big)\d\epsilon.\label{eqn:vectorperformancepredictor}
\end{align}
Note that since the probability~\eqref{eqn:multipleparameterprobability} is always larger than or equal to the probability~\eqref{eqn:singleparameterprobability}, the predicted MSE in~\eqref{eqn:vectorperformancepredictor} is always larger than or equal to the single parameter predicted MSE in~\eqref{eqn:scalarperformancepredictor}.

Although we ended up with an analytical expression for the predicted MSE in the nuisance parameter case, unfortunately, the calculation of the predicted MSE in~\eqref{eqn:vectorperformancepredictor} involves the numerical calculation of the $N_\theta$-variate normal (c)cdf which can be carried out for only small values of the number of grid points $N_\theta$. Furthermore, the covariance matrix $\Re\big\{\mathbf{\widetilde{M}}_{\epsilon}^\h\mathbf{\widetilde{M}}_{\epsilon}\big\}$ might be ill-conditioned or singular which makes the calculation of the probability even more difficult. As a result, the calculation of the predicted MSE in~\eqref{eqn:vectorperformancepredictor} would be computationally infeasible for large grid sizes $N_\theta$. To avoid this calculation we might follow an alternative approach by approximating the right hand side of~\eqref{eqn:multipleparameterderivation1} as
\begin{subequations}
    \begin{align}
        P\bigg(&\myfrac[1.5pt]{\max_{\boldsymbol{\theta}_{\backslash 1}}f(\mathbf{x}|\bar\theta_1+2\epsilon,\boldsymbol{\theta}_{\backslash 1})}{\max_{\boldsymbol{\theta}_{\backslash 1}}f(\mathbf{x}|\bar\theta_1,\boldsymbol{\theta}_{\backslash 1})}\ge 1\bigg)\nonumber\\
        &\approx\, P\bigg(\max_{\boldsymbol{\theta}_{\backslash 1}\in\boldsymbol{\Theta}_{\backslash 1}}
        \Big[2\Re\{\mathbf{\tilde{m}}^\h (\bar\theta_1+2\epsilon,\boldsymbol{\theta}_{\backslash 1};\bar\theta_1,\boldsymbol{\bar\theta}_{\backslash 1}) \mathbf{v}\}-\|\mathbf{\tilde{m}}(\bar\theta_1+2\epsilon,\boldsymbol{\theta}_{\backslash 1};\bar\theta_1,\boldsymbol{\bar\theta}_{\backslash 1})\|^2\Big]
        \ge 0\bigg)\label{eqn:PreviousApproximations}\\
        &\approx \max_{\boldsymbol{\theta}_{\backslash 1}\in\boldsymbol{\Theta}_{\backslash 1}} P\bigg(
        \Big[2\Re\{\mathbf{\tilde{m}}^\h (\bar\theta_1+2\epsilon,\boldsymbol{\theta}_{\backslash 1};\bar\theta_1,\boldsymbol{\bar\theta}_{\backslash 1}) \mathbf{v}\}-\|\mathbf{\tilde{m}}(\bar\theta_1+2\epsilon,\boldsymbol{\theta}_{\backslash 1};\bar\theta_1,\boldsymbol{\bar\theta}_{\backslash 1})\|^2\Big]
        \ge 0\bigg)\label{eqn:multipleparameterfinalapproximation}\\
        &= \max_{\boldsymbol{\theta}_{\backslash 1}\in\boldsymbol{\Theta}_{\backslash 1}} P\Big(
        2\Re\{\mathbf{\tilde{m}}^\h (\bar\theta_1+2\epsilon,\boldsymbol{\theta}_{\backslash 1};\bar\theta_1,\boldsymbol{\bar\theta}_{\backslash 1}) \mathbf{v}\}\ge \|\mathbf{\tilde{m}}(\bar\theta_1+2\epsilon,\boldsymbol{\theta}_{\backslash 1};\bar\theta_1,\boldsymbol{\bar\theta}_{\backslash 1})\|^2
        \Big)\\
        &=\max_{\boldsymbol{\theta}_{\backslash 1}\in\boldsymbol{\Theta}_{\backslash 1}}\N_{\mathrm{ccdf}}\Big(\|\mathbf{\tilde{m}}(\bar\theta_1+2\epsilon,\boldsymbol{\theta}_{\backslash 1};\bar\theta_1,\boldsymbol{\bar\theta}_{\backslash 1})\|^2;0,2\sigma^2\|\mathbf{\tilde{m}}(\bar\theta_1+2\epsilon,\boldsymbol{\theta}_{\backslash 1};\bar\theta_1,\boldsymbol{\bar\theta}_{\backslash 1})\|^2\Big)\\
        &=\max_{\boldsymbol{\theta}_{\backslash 1}\in\boldsymbol{\Theta}_{\backslash 1}}\N_{\mathrm{ccdf}}\big(\|\mathbf{\tilde{m}}(\bar\theta_1+2\epsilon,\boldsymbol{\theta}_{\backslash 1};\bar\theta_1,\boldsymbol{\bar\theta}_{\backslash 1})\|;0,2\sigma^2\big)\\
        &=\N_{\mathrm{ccdf}}\Big(\min_{\boldsymbol{\theta}_{\backslash 1}\in\boldsymbol{\Theta}_{\backslash 1}}\|\mathbf{\tilde{m}}(\bar\theta_1+2\epsilon,\boldsymbol{\theta}_{\backslash 1};\bar\theta_1,\boldsymbol{\bar\theta}_{\backslash 1})\|;0,2\sigma^2\Big),
        \label{eqn:multipleparameterderivation3}
    \end{align}
\end{subequations}
under the assumption that $\min_{\boldsymbol{\theta}_{\backslash 1}\in\boldsymbol{\Theta}_{\backslash 1}}\|\mathbf{\tilde{m}}(\bar\theta_1+2\epsilon,\boldsymbol{\theta}_{\backslash 1};\bar\theta_1,\boldsymbol{\bar\theta}_{\backslash 1})\|\neq 0$. The approximation sign in~\eqref{eqn:PreviousApproximations} represents the approximations made until reaching~\eqref{eqn:multipleparameterderivation1}. The approximation sign in~\eqref{eqn:multipleparameterfinalapproximation} can be replaced with a greater than equal to sign, i.e., the right hand side of it is a lower bound for the left hand side.     Substituting~\eqref{eqn:multipleparameterderivation3} into~\eqref{eqn:MLmultipleparametercase} gives the predicted MSE expression shown below:
\begin{align}
    \widehat{\MSE}_{\mathrm{ML}}(\bar{\theta}_1)=&\,2\int_{-\infty}^\infty|\epsilon|\N_{\mathrm{ccdf}}\Big(\min_{\boldsymbol{\theta}_{\backslash 1}\in\boldsymbol{\Theta}_{\backslash 1}}\|\mathbf{\tilde{m}}(\bar\theta_1+2\epsilon,\boldsymbol{\theta}_{\backslash 1};\bar\theta_1,\boldsymbol{\bar\theta}_{\backslash 1})\|;0,2\sigma^2\Big)\d\epsilon.\label{eqn:vectorperformancepredictor1}
\end{align}
The predicted MSE in~\eqref{eqn:vectorperformancepredictor1} is always smaller than or equal to the computationally prohibitive predicted MSE in~\eqref{eqn:vectorperformancepredictor} due to the approximation made in~\eqref{eqn:multipleparameterfinalapproximation}, however, it requires the calculation of the ccdf of only a univariate normal random variable. Note that the MSE in~\eqref{eqn:vectorperformancepredictor1} is still always larger than or equal to the single parameter MSE in~\eqref{eqn:scalarperformancepredictor} since the parameter grid $\boldsymbol{\Theta}_{\backslash 1}$ contains the true value $\boldsymbol{\bar\theta}_{\backslash 1}$ of $\boldsymbol{\theta}_{\backslash 1}$. This is because of the fact that
\begin{align}
    \min_{\boldsymbol{\theta}_{\backslash 1}\in\boldsymbol{\Theta}_{\backslash 1}}&\|\mathbf{\tilde{m}}(\bar\theta_1+2\epsilon,\boldsymbol{\theta}_{\backslash 1};\bar\theta_1,\boldsymbol{\bar\theta}_{\backslash 1})\|\le \|\mathbf{\tilde{m}}(\bar\theta_1+2\epsilon,\boldsymbol{\bar\theta}_{\backslash 1};\bar\theta_1,\boldsymbol{\bar\theta}_{\backslash 1})\|=\|\mathbf{\tilde{m}}(\bar\theta_1+2\epsilon;\bar\theta_1)\|
\end{align}
and that the function $\N_{\mathrm{ccdf}}(\cdot,0,2\sigma^2)$ monotonically increases as its argument gets smaller.

The intuitive meaning of the MSE expression~\eqref{eqn:vectorperformancepredictor1} can be explained as follows. When the nuisance parameters $\boldsymbol{\theta}_{\backslash 1}$ are known, i.e., we have the case of a single parameter in Section~\ref{sec:parametricmeansingle}, the MSE is seen to be dependent on the distance between the mean vector $\mathbf{m}(\bar\theta_1+2\epsilon,\boldsymbol{\bar\theta}_{\backslash 1})$ and the true mean vector $\mathbf{m}(\bar\theta_1,\boldsymbol{\bar\theta}_{\backslash 1})$, which was shown as (the magnitude of) the vector $\mathbf{\tilde{m}}(\bar\theta_1+2\epsilon;\bar\theta_1)\triangleq\mathbf{\tilde{m}}(\bar\theta_1+2\epsilon,\boldsymbol{\bar\theta}_{\backslash 1};\bar\theta_1,\boldsymbol{\bar\theta}_{\backslash 1})$ in~\eqref{eqn:scalarperformancepredictor}. On the other hand, when the
the nuisance parameters $\boldsymbol{\theta}_{\backslash 1}$ are not known, the predicted MSE is dependent on minimum distance between the mean vectors $\mathbf{m}(\bar\theta_1+2\epsilon,\boldsymbol{\theta}_{\backslash 1})$, where $\boldsymbol{\theta}_{\backslash 1}$ takes values in a grid containing the true nuisance parameter value $\boldsymbol{\bar\theta}_{\backslash 1}$, and the fixed true mean vector $\mathbf{m}(\bar\theta_1,\boldsymbol{\bar\theta}_{\backslash 1})$. Hence if the vector $\mathbf{m}(\bar\theta_1+2\epsilon,\boldsymbol{\theta}_{\backslash 1})$ is similar to the fixed vector $\mathbf{m}(\bar\theta_1,\boldsymbol{\bar\theta}_{\backslash 1})$ for some values of the nuisance parameter $\boldsymbol{\theta}_{\backslash 1}$ in the grid, the resulting predicted MSE would get larger.
\begin{remark}\label{rem:approximationsforgeneralcase}
The approximations made in this section on~\eqref{eqn:MLmultipleparametercase} can be applied to the general case~\eqref{eqn:actualmultipleparametercase} as follows.\begin{subequations}
\begin{align}
       P\Big(\max_{\boldsymbol{\theta}_{\backslash 1}}\L(\mathbf{x};\bar\theta_1+2\epsilon,\boldsymbol{\theta}_{\backslash 1})\ge\max_{\boldsymbol{\theta}_{\backslash 1}}\L(\mathbf{x};\bar\theta_1,\boldsymbol{\theta}_{\backslash 1})\Big)
        \approx&\, P\Big(\max_{\boldsymbol{\theta}_{\backslash 1}\in\boldsymbol{\Theta}_{\backslash 1}}\L(\mathbf{x};\bar\theta_1+2\epsilon,\boldsymbol{\theta}_{\backslash 1})\ge \L(\mathbf{x};\bar\theta_1,\boldsymbol{\bar\theta}_{\backslash 1})\Big),\\
        \approx& \max_{\boldsymbol{\theta}_{\backslash 1}\in\boldsymbol{\Theta}_{\backslash 1}} P\big(\L(\mathbf{x};\bar\theta_1+2\epsilon,\boldsymbol{\theta}_{\backslash 1})\ge\L(\mathbf{x};\bar\theta_1,\boldsymbol{\bar\theta}_{\backslash 1})\big).
\end{align}
\end{subequations}
\end{remark}

\subsection{Application to ML Estimation under Model Mismatch}\label{sec:mismatchedML}
In this section we consider the problem of ML estimation with the parametric mean model under model mismatch, also known as misspecified ML (MML) estimation in the literature~\cite{white82,Richmond2015_MCRB}. For the sake of simplicity we consider only the scalar parameter case, i.e., $\theta\in\mathbb{R}$. The measurements $\mathbf{x}$ are modeled as $\mathbf{x}=\mathbf{\bar m}(\bar\theta)+\mathbf{v}$, where $\mathbf{\bar{m}}(\cdot)$ denotes the true mean function and $\mathbf{v}\sim\CN(\mathbf{v};\mathbf{0},\bar\sigma^2\mathbf{I}_N)$ represents the measurement noise with the true variance $\bar\sigma^2$. This model corresponds to the true likelihood $\bar{f}(\mathbf{x};\bar\theta)\triangleq\CN(\mathbf{x};\mathbf{\bar m}(\bar\theta),\bar\sigma^2\mathbf{I}_N).$
We are interested in the MSE of the mismatched ML estimator $\hat{\theta}$ of $\theta$ given as
\begin{align}
\hat{\theta}\triangleq\arg\max_{\theta} f(\mathbf{x};\theta),
\end{align}
where the objective function is the assumed likelihood $f(\mathbf{x};\theta)$ given as
\begin{align}
f(\mathbf{x};\theta)\triangleq \CN(\mathbf{x};\mathbf{m}(\theta),\sigma^2\mathbf{I}).
\end{align}
For predicting the performance of the MML estimator given above, we can use the MSE expression of Remark~\ref{cor:MSE} by setting $\L(\mathbf{x};\theta)\triangleq f(\mathbf{x};\theta)$ and calculating the probability in the integrand of~\eqref{eqn:MSEHATdefinition} with respect to the true measurement distribution $\bar{f}(\cdot;\bar\theta)$. The log-likelihood ratio $\ln\frac{f(\mathbf{x};\bar\theta+2\epsilon)}{f(\mathbf{x};\bar\theta)}$ in this case is given as
\begin{align}
	 \ln&\myfrac[1.5pt]{f(\mathbf{x};\bar\theta+2\epsilon)}{f(\mathbf{x};\bar\theta)}=\frac{1}{\sigma^2}\big(2\Re\big\{\mathbf{\tilde m}^\h(\bar\theta+2\epsilon;\bar\theta) (\mathbf{x}-\mathbf{\bar m}(\bar\theta))\big\}-\|\mathbf{\tilde m}(\bar\theta+2\epsilon;\bar\theta)\|^2+2\Re\big\{\mathbf{\tilde m}^\h(\bar\theta+2\epsilon;\bar\theta)\boldsymbol\mu(\bar\theta)\big\}\big)
\end{align}
where $\mathbf{\tilde m}(\cdot,\cdot)$ was  defined in~\eqref{eqn:mtildeDefinition} and
\begin{align}
    \boldsymbol\mu(\theta)\triangleq \mathbf{\bar m}(\theta)-\mathbf{m}(\theta).\label{eqn:mudefinition}
\end{align}
We can now calculate the probability of the event $\ln({f(\mathbf{x};\bar\theta+2\epsilon)}/{f(\mathbf{x};\bar\theta)})\ge 0$ with respect to the true measurement distribution $\bar{f}(\cdot;\bar\theta)$ as
\begin{align}
    P&\bigg(\ln\myfrac[1.5pt]{f(\mathbf{x}|\bar\theta+2\epsilon)}{f(\mathbf{x}|\bar\theta)}\ge 0\bigg)=\N_{\mathrm{ccdf}}\bigg(\| \mathbf{\tilde m}(\cdot)\|-2\Re\bigg\{\frac{ \mathbf{\tilde m}^\h(\cdot)}{\| \mathbf{\tilde m}(\cdot)\|}\boldsymbol{\mu}(\bar\theta)\bigg\};0,2\bar{\sigma}^2\bigg)
\end{align}
under the assumption that $\|\mathbf{\tilde m}(\cdot)\|\neq 0$, where we dropped the arguments of the function $\mathbf{\tilde m}(\bar\theta+2\epsilon;\bar\theta)$ for brevity. Substituting this expression into the integrand of~\eqref{eqn:MSEHATdefinition} we get the following predicted MSE for the MML estimate.
\begin{align}
    \widehat{\MSE}_{\mathrm{MML}}&(\bar{\theta})=\,2\int_{-\infty}^\infty|\epsilon|\N_{\mathrm{ccdf}}\bigg(\| \mathbf{\tilde m}(\cdot)\|-2\Re\bigg\{\frac{ \mathbf{\tilde m}^\h(\cdot)}{\| \mathbf{\tilde m}(\cdot)\|}\boldsymbol{\mu}(\bar\theta)\bigg\};0,2\bar{\sigma}^2\bigg)\d\epsilon.\label{eqn:mismatchperformancepredictor}
\end{align}

\section{Numerical Results}\label{sec:numericalResults}
In this section, we examine the performance of the proposed MSE expression on four different direction of arrival  (DOA) estimation problems. The first two problems study the conventional and misspecified ML estimation respectively. In the third one, we investigate the performance of an IDE whose objective function is not the likelihood function, but a function derived from the manifold characteristics. The fourth problem investigates Bayesian DOA estimation. The implementation details of the numerical experiments are given in \ref{app:implementationDetails}.

\subsection{DOA Estimation (No Model Mismatch)} \label{sec:DOAest}
Consider the DOA estimation problem with an $N$-element sensor array with the following array manifold.
\begin{equation}
	\mathbf{a}_{\boldsymbol{\psi}} =\, [a_1, a_2, \ldots, a_N]^\t,  \quad a_n = \exp\left( j\frac{2\pi}{\lambda}\mathbf{p}_n^\t\mathbf{u}_{\boldsymbol\psi}\right), \quad
	\mathbf{u}_{\boldsymbol{\psi}} = \left[
	\begin{array}{c}
	    \cos(\phi) \sin(\theta)   \\
	    \sin(\phi) \sin(\theta)   \\
	    \cos(\theta)
	\end{array} \right], \quad \mathbf{p}_n =\left[
	\begin{array}{c}
	    p_n^{\mathsf{x}} \\
	    p_n^{\mathsf{y}} \\
	    p_n^{\mathsf{z}}
	\end{array} \right],
\end{equation}
where,  $\boldsymbol\psi\triangleq[\phi,\,\,\theta]^\t$ denotes the unknown DOA vector composed of azimuth $\phi\in[0,\,2\pi)$\,rads (measured from the x-axis in counter-clockwise direction) and elevation  $\theta\in[0,\,\pi)$\,rads (measured from the z-axis).
 $\mathbf{p}_n$ is the position vector of the $n$th sensor containing the $\mathsf{x}$, $\mathsf{y}$ and $\mathsf{z}$-coordinates;
 $N=11$ is the number of sensors;
 $\lambda$ denotes the wavelength.

\begin{table}[tb]
    \caption{Sensor positions for the array in Figure~\ref{fig:arrayConfig}} 
    \centering
    \small{
    \begin{tabular}{|c|c|c|c|c|c|c|c|c|c|c|c|}
        \hline
        Sensor-ID          & 1      & 2      & 3      & 4       & 5       & 6       & 7       & 8       & 9 & 10     & 11      \\ \hline
        x, {[}$\lambda${]} & 1.6667 & 1.1785 & 0 & -1.1785 & -1.6667 & -1.1785 & 0       & 1.1785  & 0 & 0      & 0       \\ \hline
        y, {[}$\lambda${]} & 0      & 1.1785 & 1.6667 & 1.1785  & 0       & -1.1785 & -1.6667 & -1.1785 & 0 & 0      & 0       \\ \hline
        z, {[}$\lambda${]} & 0      & 1.1785 & 1.6667 & 1.1785  & 0       & -1.1785 & -1.6667 & -1.1785 & 0 & 1.6667 & -1.6667 \\ \hline
    \end{tabular}}
    \label{tab:sensorPos}
\end{table}
\begin{figure}[tb]
	\centering
	\subfloat[Array configuration.]{\includegraphics[height=0.35\columnwidth]{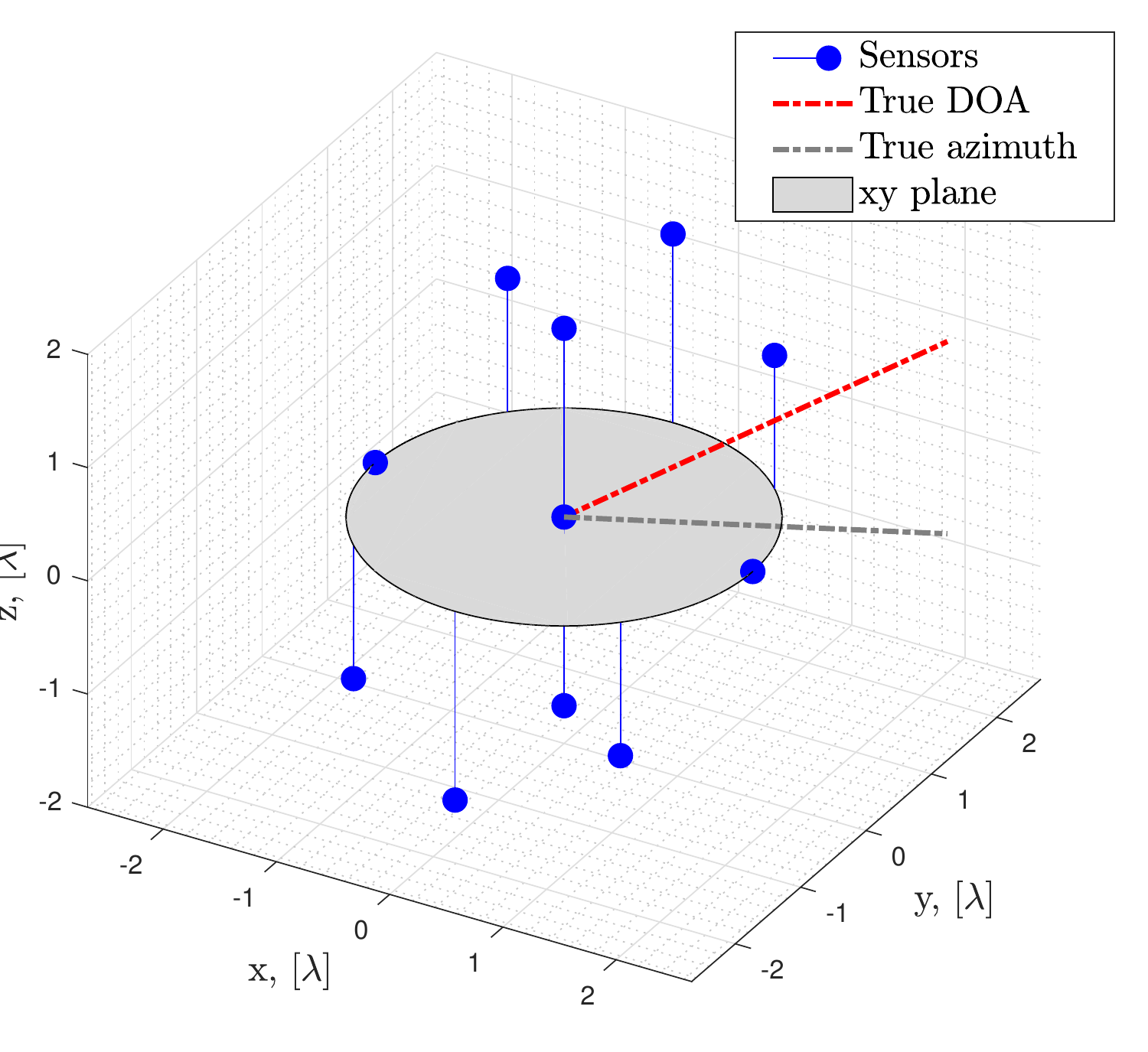}%
	\label{fig:arrayConfig}}
	\subfloat[Array beampattern at true DOA, $\phi=25^\circ$, $\theta=60^\circ$.]{\includegraphics[width=0.45\columnwidth]{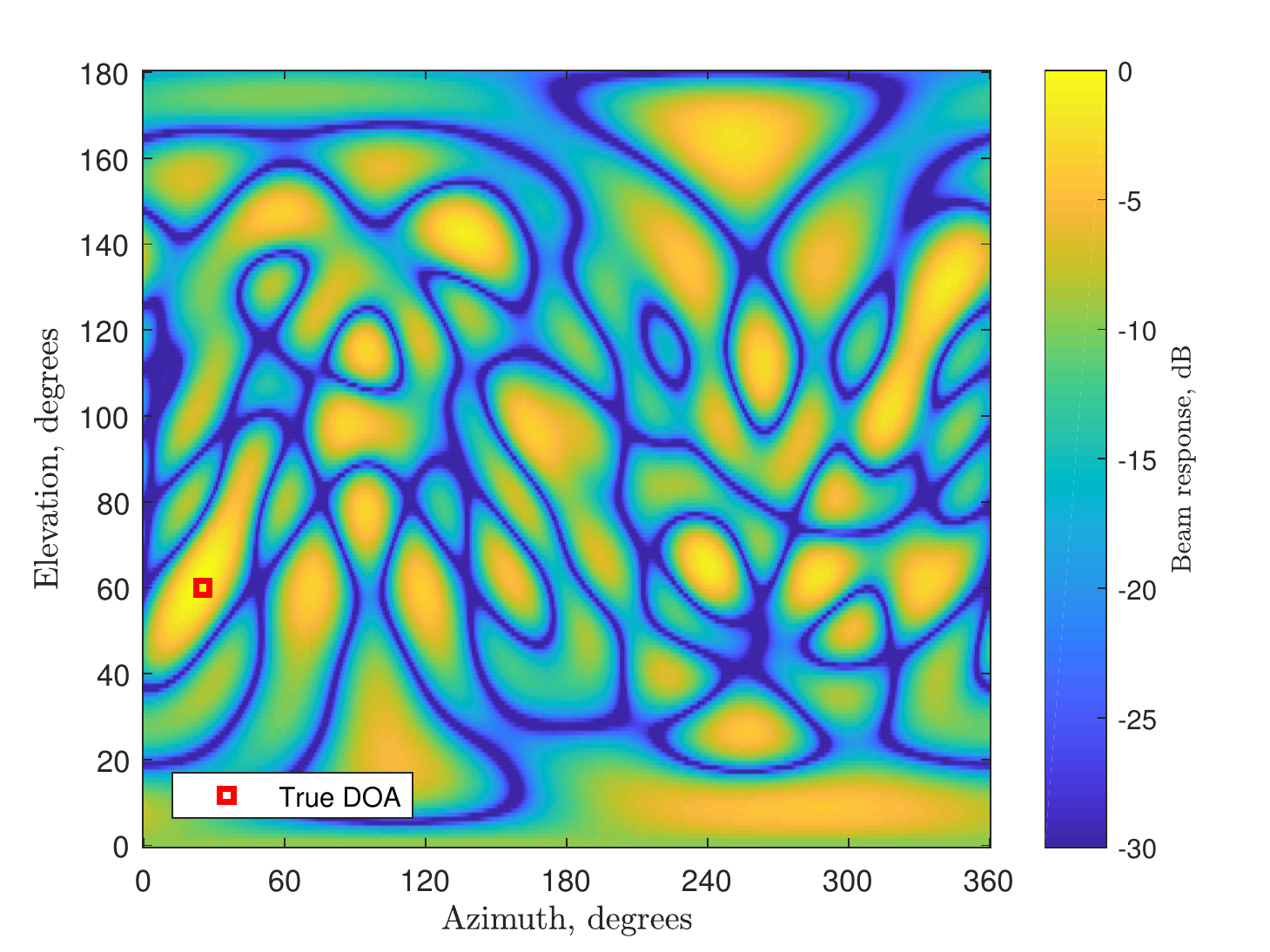}%
	\label{fig:3D_BeamPattern}}
	\caption{Array configuration and array beampattern at true DOA.}
	\label{fig:ArrayConfigAndBeampattern}
\end{figure}
\begin{figure}[tb]
	\centering
	\subfloat[Azimuth estimation at the true DOA: $\bar\phi=25^\circ $, $\bar\theta=60^\circ$.]{\includegraphics[width=0.475\columnwidth]{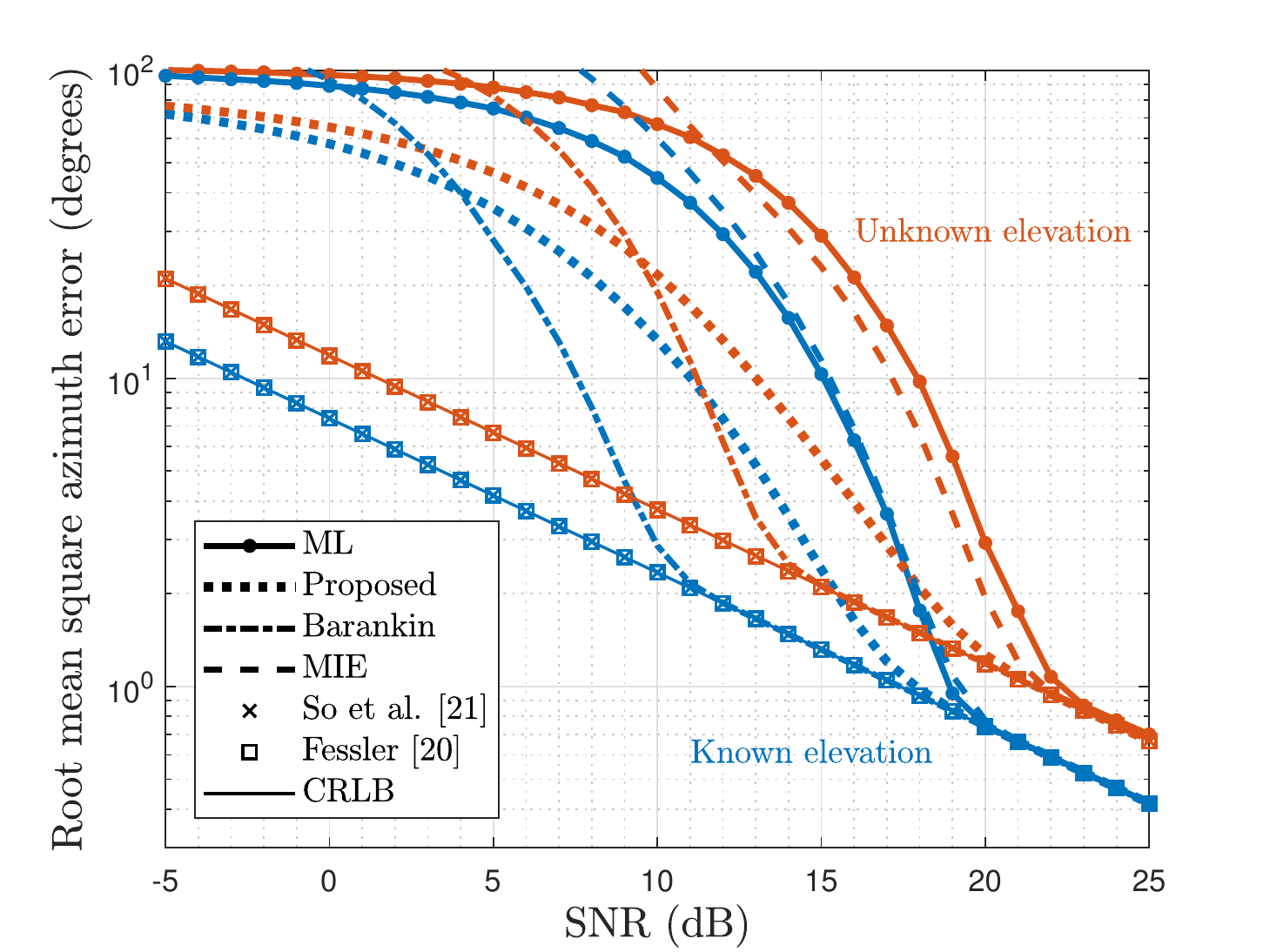}%
		\label{fig:azimuthEstimation}}
	\quad
	\subfloat[Elevation estimation at the true DOA: $\bar\phi=25^\circ $, $\bar\theta=60^\circ$.]{\includegraphics[width=0.475\columnwidth]{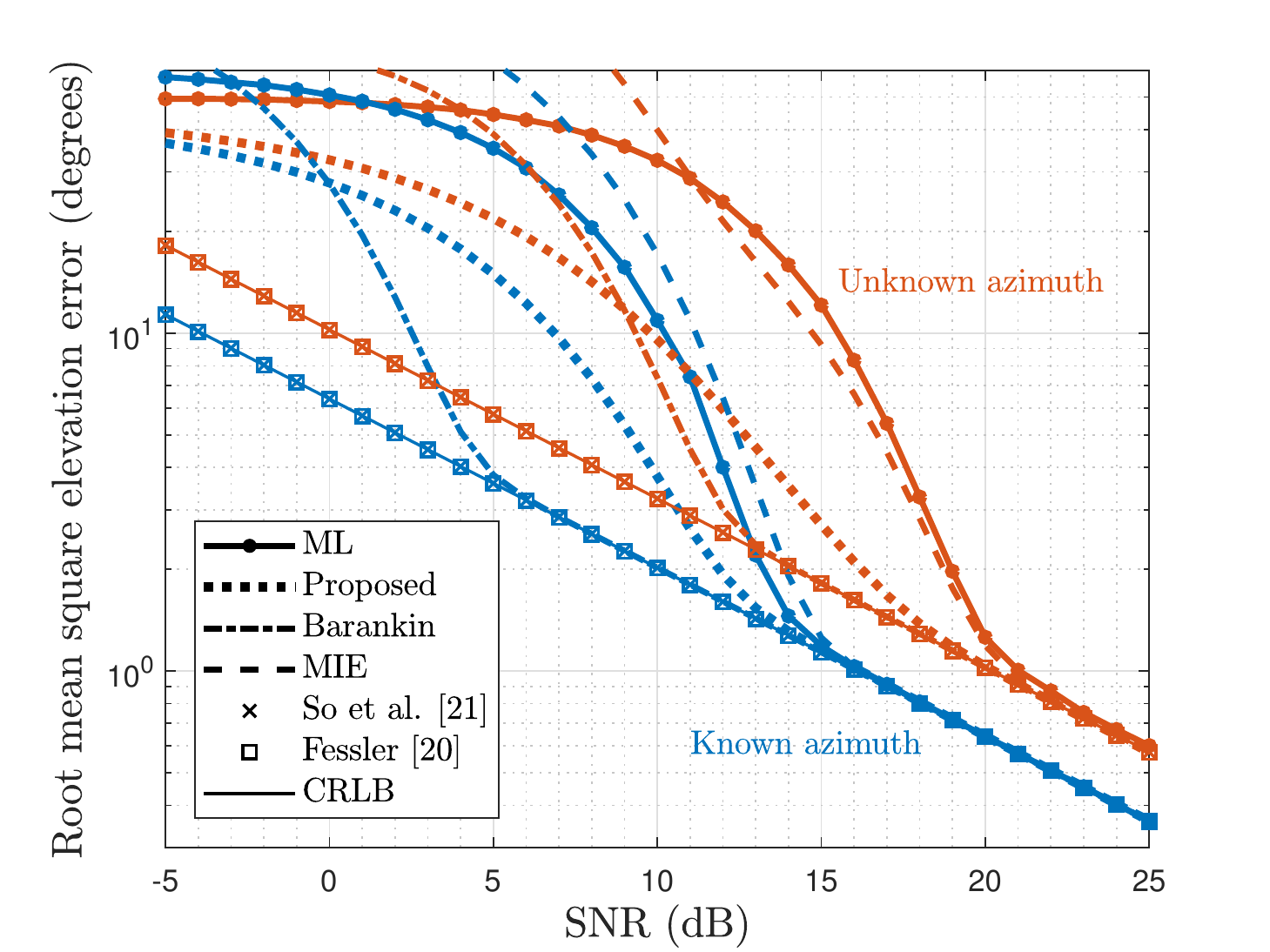}%
		\label{fig:elevationEstimation}}
	\caption{Azimuth and elevation estimation performance curves for the array configuration in Figure~\ref{fig:arrayConfig}. Blue and orange colors correspond to the cases without and with the nuisance parameter, respectively.}
	\label{fig:DOAestimation}
\end{figure}
The array, whose sensor positions are given in Table~\ref{tab:sensorPos}, is illustrated in Figure~\ref{fig:arrayConfig}.
The sensor measurement vector $\mathbf{x}\in\mathbb{C}^{N}$ under additive  noise is modeled as
\begin{equation}
	\mathbf{x} = \beta \mathbf{a}_{\boldsymbol{\bar{\psi}}} + \mathbf{v}, \label{eq:DOAmodel}
\end{equation}
 where $\mathbf{v}\sim \CN(\mathbf{v};\mathbf{0}, \sigma^2 \mathbf{I}_N)$; $\beta\in \mathbb{R}$ ($\beta$ is taken as a real-valued scalar with no loss of generality due to the circular symmetry of the complex Gaussian noise) and $\boldsymbol{\bar{\psi}}\triangleq[\bar{\phi},\,\,\bar\theta]^\t$ denotes the true value of the angle vector $\boldsymbol\psi$.
The true target angular positions are  $\bar\phi=25^\circ$ and $\bar\theta=60^\circ$. The beampattern of the array, obtained using the conventional, i.e., Bartlett, beamformer with coefficients steered to the true DOA, for this angular position is shown in Figure~\ref{fig:3D_BeamPattern}. The beampattern contains sidelobes as high as $-2$ dB, with a response normalized to $0$ dB at the true DOA. Consequently the array is prone to gross errors. With these definitions, the ML estimator involves the following optimization problem:
\begin{align}
	\boldsymbol{\hat{\psi}} =& \arg\max_{\boldsymbol\psi} \Re\{\mathbf{x}^\h \mathbf{a}_{\boldsymbol\psi}\} \label{eq:MLforVectorDOA}.
\end{align}
Since the signal model is a parametric mean model with the mean function $\mathbf{m}(\boldsymbol{\psi}) = \beta \boldsymbol{a_{\psi}}$, the finite support versions of the expressions \eqref{eqn:scalarperformancepredictor} and \eqref{eqn:vectorperformancepredictor1} can be utilized for MSE prediction. We consider three different cases: (i) Azimuth $\phi$ is unknown, but elevation $\theta=\bar\theta$ is known; (ii) Elevation $\theta$ is unknown, but azimuth $\phi=\bar\phi$ is known; (iii) Both azimuth $\phi$ and elevation $\theta$ are unknown.
%
The results of $10^5$ Monte Carlo simulations are given in Figure~\ref{fig:azimuthEstimation} and Figure~\ref{fig:elevationEstimation} for azimuth and elevation estimates, respectively. For comparison purposes the corresponding CRLBs (see \cite{CRB_3Darray} for the analytical expressions), Barankin bounds (BBs)~\cite{BarankinBound1971} with single test point optimized over a grid, Fessler's method~\cite{Fessler1996}, So et al.'s method~\cite{SimpleFormulasForBiasAndMSE}, and method of interval errors (MIE)~\cite{Athley_thresholdSNR_2005} are also illustrated. As seen from Figure~\ref{fig:DOAestimation}, the proposed method is able to predict the threshold $\SNR$ below which the ML estimator starts following the CRLB and tracks the CRLB in the asymptotic region as expected. BB, on the other hand, converges to CRLB at a much smaller SNR value than the ML estimator. MIE closely follows the ML estimator in the threshold region. This is essentially due to the problem specific selection of the intervals and accurate gross error probability calculation. Note that MIE does not have any assumptions on the objective function, such as symmetry or unimodality, leading to a better tracking of ML estimator performance especially in the threshold region. Taylor expansion based methods of Fessler and So et al.\ follow the CRLB values in all regions of operation and they are unable to take into account the gross errors the ML estimator makes below the threshold SNR.

\subsection{DOA Estimation (Model Mismatch)} \label{sec:DOAest_modelMismatch} In this section we consider the misspecified ML estimation problem examined in Section~\ref{sec:mismatchedML} on the parameterized mean model. For this purpose  we consider the near field azimuth estimation problem with known elevation angle, in which the estimator uses the plane wave propagation assumption (far field assumption) rather than the true propagation model which is the spherical spreading.

\begin{figure}[tb]
    \centering
    \begin{minipage}{.475\textwidth}
        \centering
    	\includegraphics[width=0.95\columnwidth]{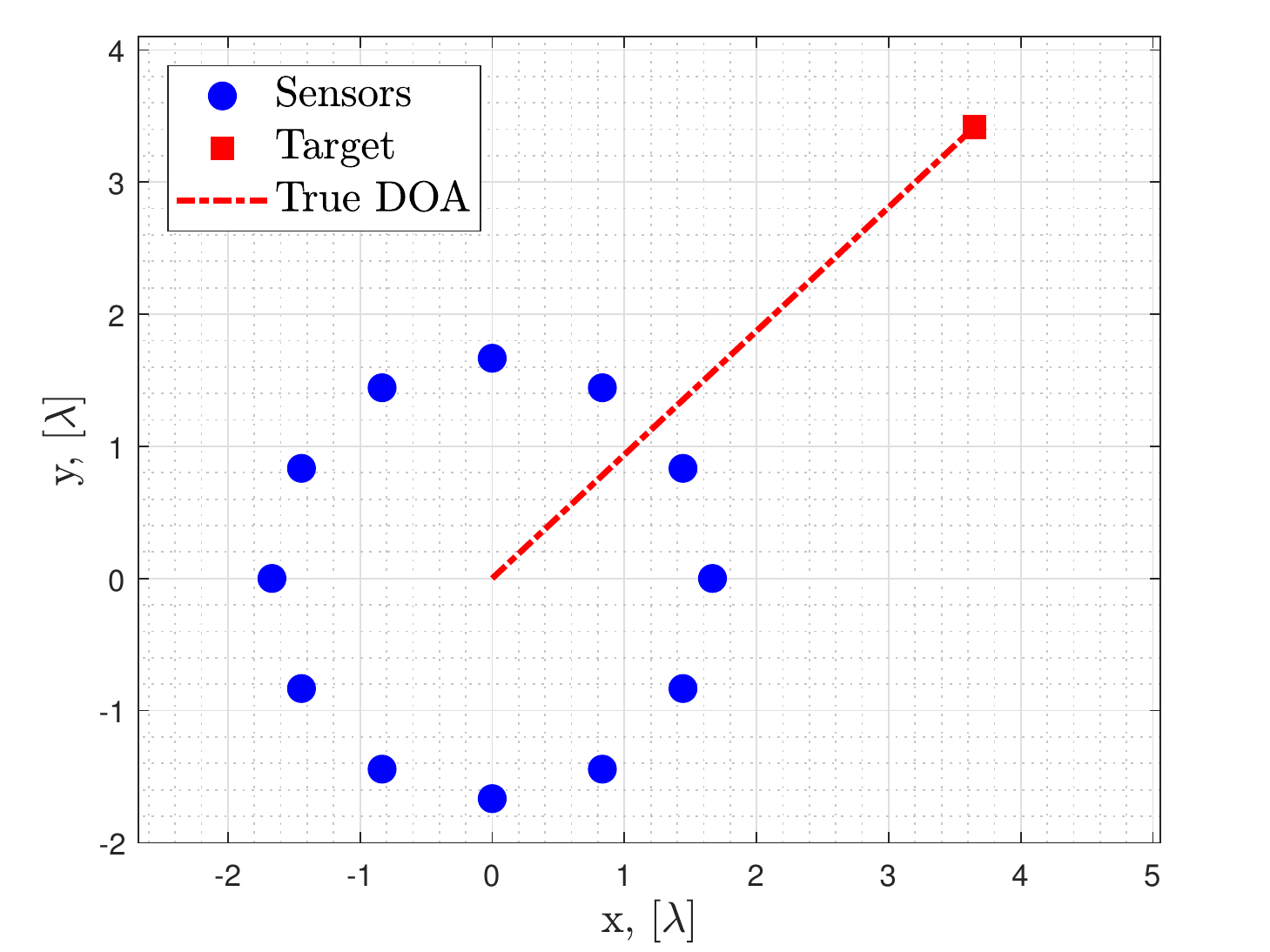}
    	\caption{12 element uniform circular array with a radius of $\frac{5}{3}\lambda$ and target of interest at $5\lambda$ range.}
    	\label{fig:neafield_7L}
    \end{minipage}
    \qquad
	\begin{minipage}{.475\textwidth}
	    \centering
		\includegraphics[width=0.975\columnwidth]{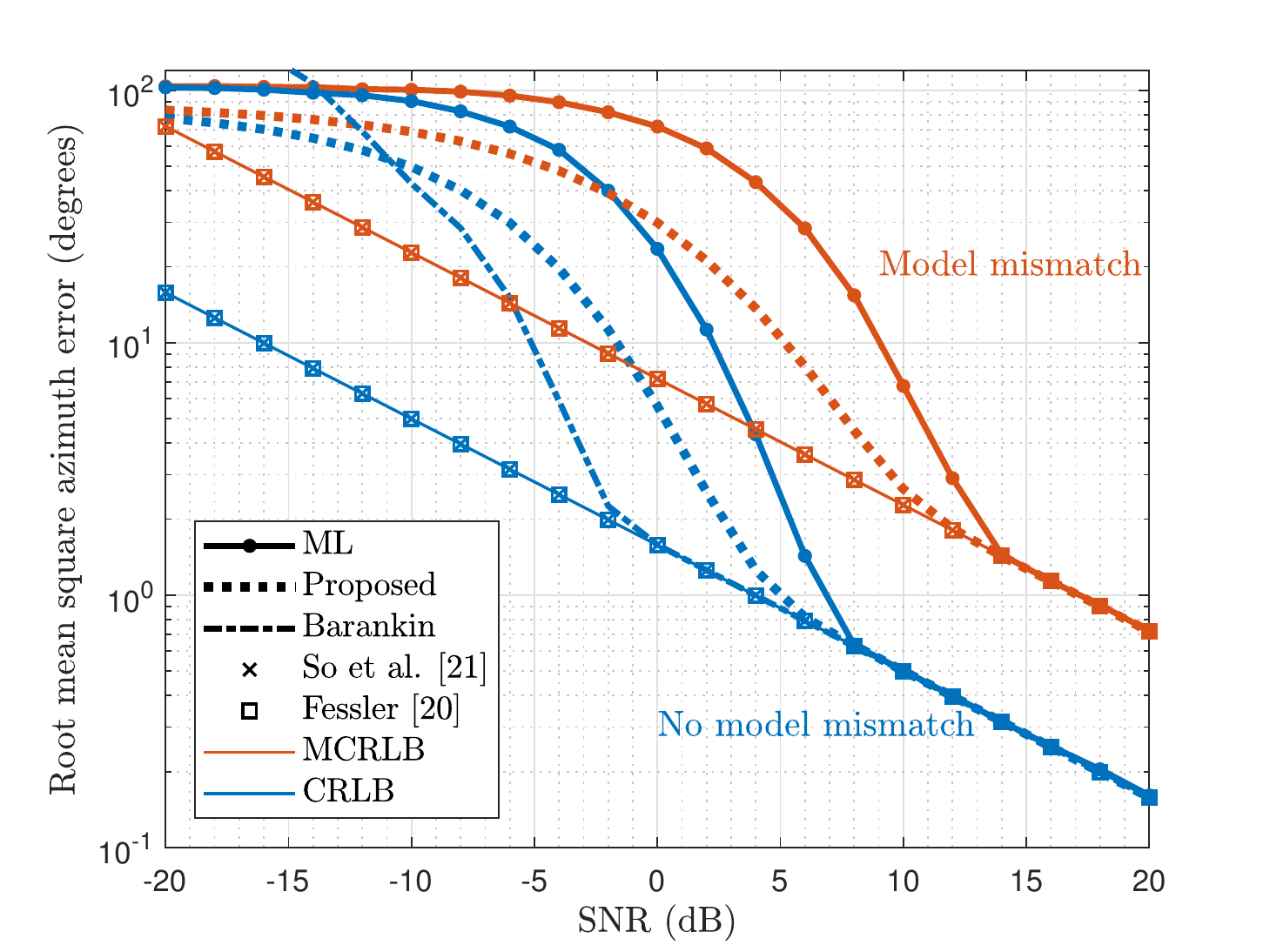}
    	\caption{Near-field and far-field performance of a 12 element uniform circular array, for a target at $5\lambda$ distance.}
    	\label{fig:mismatch_5L}
	\end{minipage}
\end{figure}

A uniform circular array of radius $5\lambda/3$ with 12 elements is used. The signal of interest emanates from a target at a range of $5\lambda$, which is closer than the far-field limit $2(10\lambda/3)^2 / \lambda = 200\lambda/9$ \cite{vanTreesOAP}. The array configuration and the target position are illustrated in Figure~\ref{fig:neafield_7L}.

The true signal model is given as $\mathbf{\bar{m}}(\bar\phi) =\, [\bar{a}_1, \bar{a}_2, \ldots, \bar{a}_N]^\t, \bar{a}_n = \exp\big(-j\frac{2\pi}{\lambda} d_n(\bar\phi)\big),$ where
$d_n(\phi) = \, \|\mathbf{p}_n-r\mathbf{u}_{\phi}\|$,
$\mathbf{u}_{\phi} = \, [ \cos(\phi),\, \sin(\phi) ]^\t$,
$\mathbf{p}_n = \, [ p_n^\mathsf{x},\, p_n^\mathsf{y} ]^\t$
and $r$ is the range of the target from the array center as illustrated in Figure~\ref{fig:neafield_7L}.
The assumed model by the estimator is the plane wave model, given as
$\mathbf{m}(\phi) = \, [a_1, a_2, \ldots, a_N]^\t$,
$a_n = \exp\big( j\frac{2\pi}{\lambda}\mathbf{p}_n^\t\mathbf{u}_\phi\big).
$
There is no misspecification in the noise variance, i.e., $\sigma^2=\bar{\sigma}^2$.

The MSE values for this experiment with 10,000 Monte Carlo runs are given in Figure~\ref{fig:mismatch_5L} along with the corresponding MCRLBs~\cite{FortunatiGGR2017,Richmond2015_MCRB}, BBs~\cite{BarankinBound1971}, and the results for Fessler's \cite{Fessler1996} and So et al.'s \cite{SimpleFormulasForBiasAndMSE} methods. MCRLB reduces to the following expression for this specific problem.
	\begin{align}
		\MCRB(\bar\phi) =
		\CD^{-1}_D(\bar\phi) \I_D(\bar\phi) \CD^{-1}_D(\bar\phi),
	 \end{align}
where
	\begin{align}
		\I_D(\phi) &= \frac{2}{\sigma^2}\bigg\|\frac{\partial \mathbf{m}(\phi)}{\partial \phi}\bigg\|^2, \quad \CD_D(\theta) = -\I_D(\phi)+2\Re\bigg\{\bigg[\frac{\partial^2 \mathbf{m}(\phi)}{\partial \theta^2} \bigg]^\h \boldsymbol{\mu}(\phi)\bigg\},
	\end{align}
and $\boldsymbol{\mu}(\cdot)$ was defined in~\eqref{eqn:mudefinition}.
 BB~\cite{BarankinBound1971} with a single test point optimized over a grid, which is also the HCRB~\cite{ChapmanRobbinsBound, ForsterLarzabal2002} can be  expressed as follows.
 \begin{align}
 	\BB(\bar\phi)= \HCRB(\bar\phi) = \max_{\phi} \frac{(\phi - \bar\phi)^2}{e^{\frac{2}{\sigma^2}\|\mathbf{m}(\phi)-\mathbf{m}(\bar\phi)\|^2}-1}. \label{eq:HCR}
 \end{align}
The results given in Figure~\ref{fig:mismatch_5L} indicate that the proposed MSE expression again predicts the threshold SNR quite closely and tracks MCRLB in the small error region. On the other hand BB is optimistic about the threshold SNR and both Fessler's and So et al.'s methods yield the same results as CRLB and MCRLB for the no model mismatch and model mismatch cases respectively.

	\subsection{DOA Estimation by an IDE (ESPRIT)}
	\label{IDEexample}
	We consider a uniform linear array composed of $N=15$ sensors with $\lambda/2$ element spacing. The signal model  is as follows
	\begin{align}
		x_n &= \underbrace{\alpha e^{j \pi \cos(
	\bar\phi) n}}_{\triangleq m_n(\bar\phi)} + w_n,
		\quad n=0,1,\ldots,N-1,
	\label{mndef}	
	\end{align}
	where $\alpha \in \mathbb{C}$ is the unknown complex amplitude, $w_n \sim \CN(w_n,0, \sigma_w^2)$, and $\bar\phi=35\pi/180$\,rad is the unknown true azimuth angle to be estimated. We denote the spatial frequency with $\bar\omega$ and define
	$
	    \bar\omega \triangleq \pi\cos(\bar\phi).
	$
	
\begin{figure}[tb]
    \centering
    \begin{minipage}{.475\textwidth}
        \centering
    	\includegraphics[width=0.975\columnwidth]{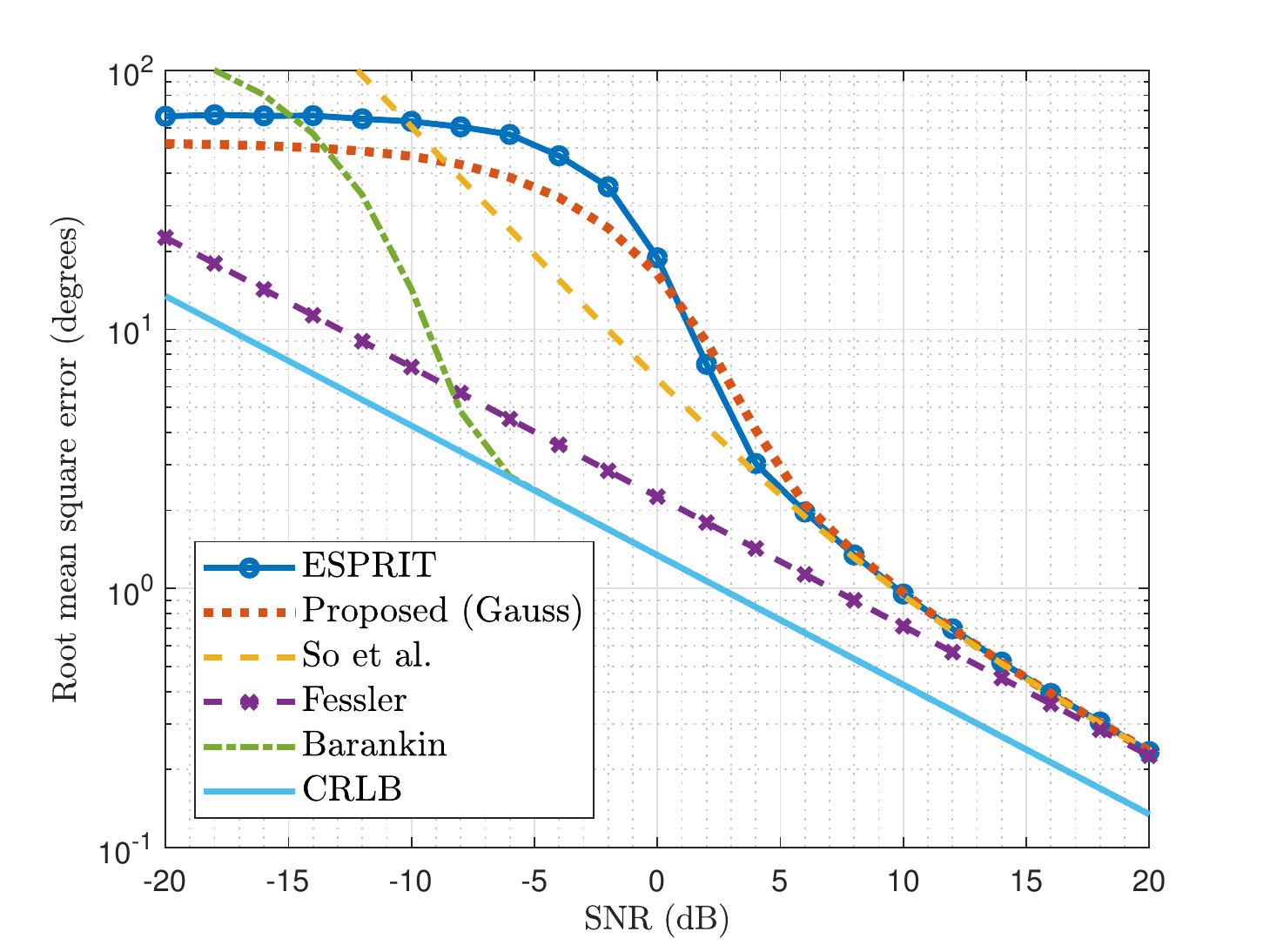}
        \caption{Non-random DOA estimation performance of ESPRIT along with values of different bounds and MSE prediction expressions.}
        \label{fig:ESPRIT}
    \end{minipage}
    \qquad
	\begin{minipage}{.475\textwidth}
	    \centering
		\includegraphics[width=0.975\columnwidth]{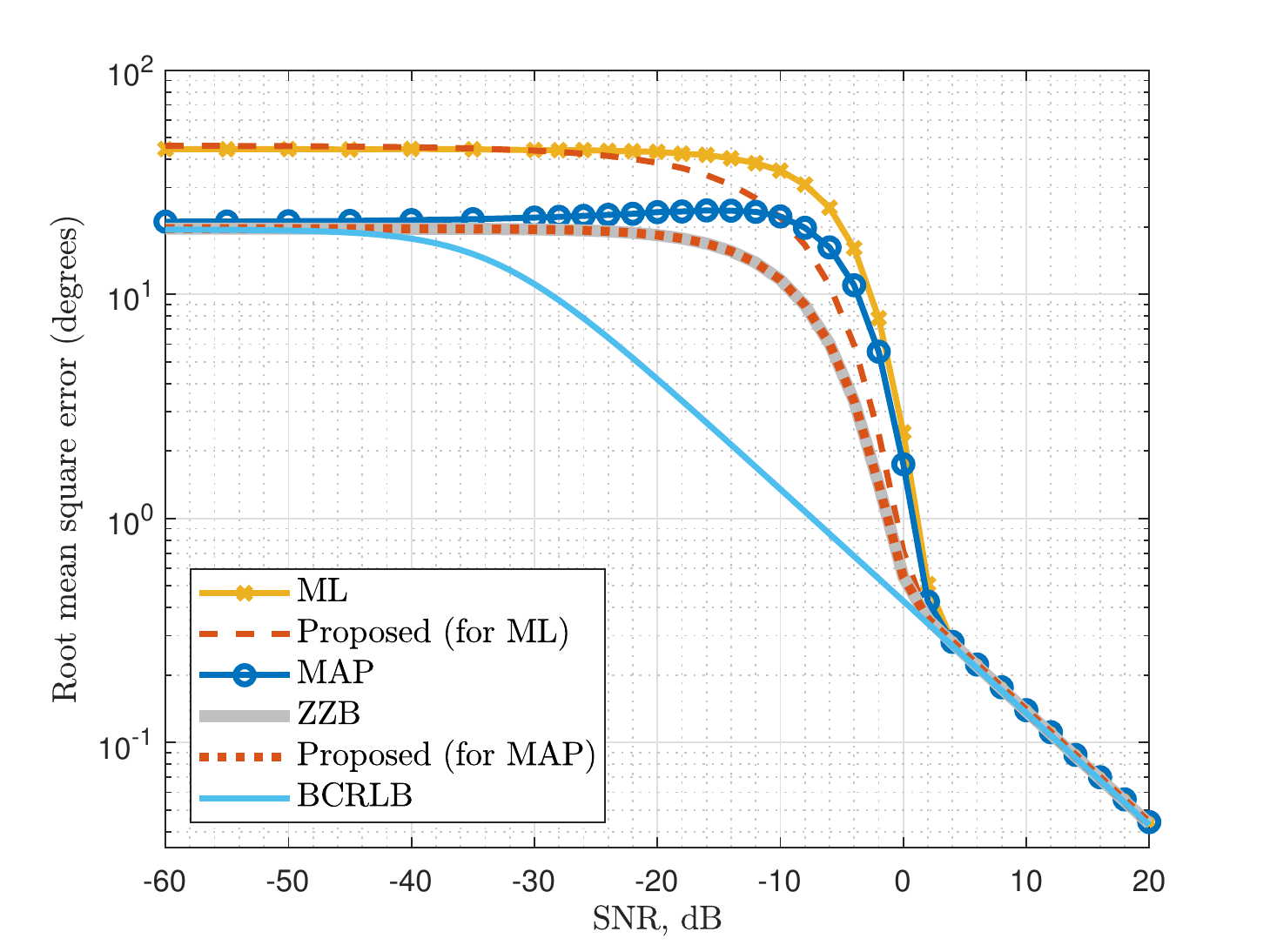}
		\caption{Bayesian DOA estimation performance of MAP and ML estimators along with the values of BCRLB, ZZB and the proposed MSE prediction expressions (for ML and MAP). }
		\label{fig:BayesianComparison}
	\end{minipage}
\end{figure}

	Due to the structure of uniform linear arrays, we can write
	$
	    m_n(\bar\phi) = e^{j\bar\omega}m_{n-1}(\bar\phi)
	$
	for the elements of the array manifold vector $m_n(\bar\phi)$ in \eqref{mndef}, which is the rotational invariance property exploited in ESPRIT~\cite{RoyKailathESPRIT1989}. Using this property we can define a somewhat adhoc cost function as follows
	\begin{align}
	    J(\omega)=\sum_{n=1}^{N-1} |x_n-e^{j\omega}x_{n-1}|^2. \label{eq:EspritCost}
	\end{align}
	By minimizing \eqref{eq:EspritCost}, we can get an estimate for $\omega$ as $\hat\omega\triangleq \arg\min_{\omega} J(\omega)=\arg\big(\sum_{n=1}^{N-1}x_{n-1}^* x_{n} \big)$; from which an estimate for the DOA can be generated as $\hat{\phi}\triangleq \arccos\big(\frac{\hat\omega}{\pi}\big)$, which we call the ESPRIT estimate. Note that the cost function $J(\cdot)$  is neither symmetric around the estimate, nor is unimodal. Hence it does not satisfy the conditions for which the proposed method yields the true MSE. The cost function $J(\omega)$ in~\eqref{eq:EspritCost} can be written in matrix form as follows.
	\begin{equation}
	        J(\omega) = \left\| \mathbf{A}_1 \mathbf{x} - e^{j\omega}\mathbf{A}_0 \mathbf{x}\right\|^2 = \mathbf{x}^\h \left( \mathbf{A}_1 - e^{j\omega}\mathbf{A}_0\right)^\h \left( \mathbf{A}_1 - e^{j\omega}\mathbf{A}_0\right) \mathbf{x}, \label{eq:Jw}
	\end{equation}
	where
    \begin{equation}
        \mathbf{x} = \left[ \begin{array} {cccc} x_0 & x_1 & \ldots & x_{N-1}\end{array}\right]^\t, \quad \mathbf{A}_0 = \left[ \begin{array}{cc} \mathbf{I}_{(N-1)} & \mathbf{0}_{(N-1) \times 1} \end{array}\right], \quad
	    \mathbf{A}_1 = \left[ \begin{array}{cc} \mathbf{0}_{(N-1) \times 1} & \mathbf{I}_{(N-1)} \end{array}\right].
    \end{equation}
    Using $\omega\triangleq\pi\cos(\phi)$,  we get,
	\begin{equation}
	    J(\phi) = \mathbf{x}^\h \big( \mathbf{A}_1 - e^{j\pi \cos(\phi)}\mathbf{A}_0\big)^\h \big( \mathbf{A}_1 - e^{j\pi \cos(\phi)}\mathbf{A}_0\big) \mathbf{x}. \label{eq:J}
	\end{equation}
	Note that in order to use the approximate MSE expression in \eqref{eqn:MSEHATdefinition}, we need to evaluate the following probability,
	\begin{align}
        P \left( J(\bar{\phi} + 2\epsilon)\le J(\bar{\phi}) \right) &= P \left( J(\bar{\phi} + 2\epsilon) - J(\bar{\phi}) \le 0\right) \triangleq P(\Delta J_{2\epsilon} \le 0), \label{eqn:DeltaJprobabilities}
    \end{align}
	where the inequalities are the reverse of those in Remark~\ref{cor:MSE} since we have a minimization problem instead of a maximization problem in our IDE.
	Using \eqref{eq:J} and after some basic algebraic operations we can express $\Delta J_{2\epsilon}\triangleq J(\bar{\phi} + 2\epsilon) - J(\bar{\phi})$ as
	$\Delta J_{2\epsilon}
	=	\mathbf{x}^\h \mathbf{Q} \mathbf{x}$ where
	\begin{align}
	    \mathbf{Q} &\triangleq (e^{j\pi\cos(\bar\phi)}-e^{j\pi\cos(\bar\phi+2\epsilon)})\mathbf{A}_1^\h \mathbf{A}_0 + (e^{-j\pi\cos(\bar\phi)}-e^{-j\pi\cos(\bar\phi+2\epsilon)})\mathbf{A}_0^\h \mathbf{A}_1.
	\end{align}
Even though the density of the quadratic form
$\Delta J_{2\epsilon} =	\mathbf{x}^\h \mathbf{Q} \mathbf{x}$
is known to be the generalized chi-squared distribution and
can be evaluated numerically \cite[Appendix A]{LinearModelsTimeSeries-Book}, we pursue a Gaussian fit to the density in order to simplify the probability calculations. To do that, we evaluate the first two moments of $\Delta J_{2\epsilon}$. Using the fact that $\mathbf{x}^\h\mathbf{Q}\mathbf{x}$ is always real, we can reach the following expressions (after some algebra)
\begin{subequations}	
	\begin{align}
	    \mu_\Delta(\bar{\phi},\epsilon) =&\, \E\{\Delta J_{2\epsilon}\} =\,\sigma_w^2\tr\big(\mathbf{Q})+\mathbf{m}^\h(\bar\phi)\mathbf{Q}\mathbf{m}(\bar\phi),\\
	    \sigma^2_\Delta(\bar{\phi},\epsilon) =&\,\mathrm{Var}\{\Delta J_{2\epsilon}\}  =\,\sigma_w^4\tr(\mathbf{Q}^2)+2\sigma_w^2\mathbf{m}^\h(\bar\phi)\mathbf{Q}^2\mathbf{m}(\bar\phi),
	\end{align}
\end{subequations}
where $\mathbf{m}(\phi)\triangleq[m_0(\phi),\,m_1(\phi),\cdots, m_{N-1}(\phi)]^\t$ and we used the result
$
    \E[(\mathbf{y}^\h\mathbf{Q}\mathbf{y})^2]=\tr^2(\mathbf{Q}\boldsymbol\Sigma)+\tr((\mathbf{Q}\boldsymbol\Sigma)^2)
$
for any Hermitian matrix $\mathbf{Q}$ and $\mathbf{y}\sim\CN(\mathbf{y};\mathbf{0},\boldsymbol\Sigma)$~\cite[Ch. V, Lemma 2.2]{miller1974complex}. With the Gaussian fit, an approximation to the suggested MSE expression becomes
\begin{equation}
    \widehat{\MSE}(\bar\phi) \approx 2\int_{-\frac{\bar{\phi}}{2}}^{\frac{\pi-\bar{\phi}}{2}} |\epsilon| \N_{\mathrm{cdf}} \left(0; \mu_\Delta(\bar{\phi},\epsilon), \sigma^2_{\Delta}(\bar{\phi},\epsilon)\right) \mathrm{d}\epsilon.
\end{equation}
where we used~\eqref{eqn:finiteSupport} with $\phi_{\min}=0$; $\phi_{\max}=\pi$ to set the integration limits and the cdf of the normal distribution is used instead of the ccdf due to the reversal of the inequalities in~\eqref{eqn:DeltaJprobabilities}. Figure~\ref{fig:ESPRIT} shows the results of 10,000 Monte Carlo runs for this experiment. The CRLB, BB with single test point optimized over a grid,  Fessler's \cite{Fessler1996} and So et al.'s \cite{SimpleFormulasForBiasAndMSE} methods are also illustrated for comparison purposes. Note that the estimator in this experiment is not efficient, hence its performance does not reach CRLB at high SNR.
 Consequently, the estimator performance is not characterized by the CRLB in any SNR region. Therefore, one needs the asymptotic MSE values as well as the pairwise error probabilities in order to calculate the MSE prediction using MIE. BB provides a very optimistic prediction for this specific problem as in the earlier examples. Although Fessler's and So et al.'s methods predicted the estimator performance well at high SNR region, they have difficulty in representing gross errors of the estimator for low SNR values. The proposed method, on the other hand, closely follows the estimator performance in all SNR regions.

\subsection{Bayesian DOA Estimation}\label{sec:bayesianDOA}
We consider the DOA estimation problem in Section~\ref{IDEexample} in a Bayesian framework. The unknown angle $\phi$ has now a prior density $f(\phi)$, which is given as the symmetric beta distribution
\begin{equation}
    f(\phi)=\frac{1}{\pi\beta(a, a)}\left(\frac{\phi}{\pi}\right)^{a-1}\left(\frac{\pi-\phi}{\pi}\right)^{a-1}, \quad\beta(a,b) \triangleq\int_{0}^{1}\phi^{a-1}(1-\phi)^{b-1}\mathrm{d}\phi, \quad 0\le \phi \le \pi
\end{equation}
with $a=10$, and the performance of the ML and MAP estimators is examined.
The proposed Bayesian MSE expression for the ML estimator can be expressed as $\widehat{\MSE}_{\mathrm{ML}} = \int_{0}^{\pi} f(\phi) \widehat{\MSE}_{\mathrm{ML}}(\phi) \mathrm{d}\phi
$ where
\begin{align}
\widehat{\MSE}_{\mathrm{ML}}(\phi) = 2 \int_{\frac{-\phi}{2}}^{\frac{\pi-\phi}{2}} |\epsilon| \mathcal{N}_{\mathrm{ccdf}} \left( \| \mathbf{\tilde{m}}(\phi+2\epsilon; \phi)\|; 0, 2\sigma_w^2\right) \mathrm{d}\epsilon,
\label{MSEMl}
\end{align}
where the integration limits are selected as in~\eqref{eqn:finiteSupport} with $\phi_{\mathrm{min}}\triangleq 0$ and $\phi_{\mathrm{max}}\triangleq\pi$. The proposed MSE expression for the MAP estimator can be expressed as $
\widehat{\MSE}_{\mathrm{MAP}} = \int_{0}^{\pi} f(\phi) \widehat{\MSE}_{\mathrm{MAP}}(\phi) \mathrm{d}\phi
$ where
\begin{align}
    \widehat{\MSE}_{\mathrm{MAP}}(\phi) = 2 \int_{-\pi}^{\pi} |\epsilon| \mathcal{N}_{\mathrm{ccdf}} \left( \| \mathbf{\tilde{m}}(\phi+2\epsilon; \phi)\| + \frac{\sigma_w^2}{\| \mathbf{\tilde{m}}(\phi+2\epsilon; \phi)\|} \log\left( \frac{f(\phi)}{f(\phi+2\epsilon)} \right); 0, 2\sigma_w^2\right) \mathrm{d}\epsilon.
\label{MSEMap}
\end{align}
Note that $\widehat{\MSE}_{\mathrm{MAP}}(\phi)$ in \eqref{MSEMap} reduces to $\widehat{\MSE}_{\mathrm{ML}}(\phi)$ in \eqref{MSEMl} when the prior is flat. BCRLB for this Bayesian estimation problem is given as \cite{VanTrees_BayesianBounds}
\begin{align}
    \mathrm{BCRLB}=\bigg(\pi^2\mathrm{SNR}\frac{N(N-1)(2N-1)}{3}\int_0^\pi\sin^2\phi f(\phi)\d\phi+\frac{4(a - 1)(2a - 1)}{\pi^2(a - 2)}\bigg)^{-1},
\end{align}
where $\SNR\triangleq \frac{|\alpha|^2}{\sigma_w^2}$ and $N=15$. ZZB (without the valley filling function) for the problem can be expressed as
\begin{align}
\mathrm{ZZB} = \frac{1}{2}\int_{0}^{\pi} \int_0^{\pi} h(f(\phi)+f(\phi+h))P_{\mathrm{min}}^{e}(\phi, \phi+h)\mathrm{d}\theta \mathrm{d}h,
\end{align}
where the minimum error probability $P_{\mathrm{min}}^{e}(\phi_1, \phi_2)$  can be calculated as
\begin{align}
    P_{\mathrm{min}}^{e}(\phi_1, \phi_2) =& \, \pi_1 \mathcal{N}_{\mathrm{ccdf}} \left( \| \mathbf{\tilde{m}}(\phi_2; \phi_1)\| + \frac{\sigma_w^2}{\| \mathbf{\tilde{m}}(\phi_2; \phi_1)\|} \log\frac{\pi_1}{\pi_2}; 0, 2\sigma_w^2 \right) \nonumber \\
    & \quad + \pi_2 \mathcal{N}_{\mathrm{ccdf}} \left( \| \mathbf{\tilde{m}}(\phi_1; \phi_2)\| + \frac{\sigma_w^2}{\| \mathbf{\tilde{m}}(\phi_1; \phi_2)\|} \log\frac{\pi_2}{\pi_1}; 0, 2\sigma_w^2 \right),
\end{align}
with the prior probabilities $\pi_1 \triangleq \frac{f(\phi_1)}{f(\phi_1) + f(\phi_2)}$ and $\pi_2 \triangleq  1 - \pi_1$.

Figure~\ref{fig:BayesianComparison} shows the RMSE performances of the MAP and ML estimators over 10,000 Monte Carlo runs for each $\SNR$ value along with the values of BCRLB, ZZB and the proposed MSE prediction expressions $\widehat{\MSE}_{\mathrm{ML}}$ and $\widehat{\MSE}_{\mathrm{MAP}}$. The values of $\mathrm{ZZB}$ and the proposed $\MSE$ prediction expression $\widehat{\MSE}_{\mathrm{MAP}}$ are identical, as expected from the results of  Section~\ref{sec:relationtoZZB}.

\section{Conclusions}
In this study we propose an MSE expression for the performance prediction of IDEs of non-random parameters. The method provides the exact MSE value when the objective function of the IDE is unimodal and symmetric. Even though, this is a rather stringent restriction for the general practice;  the symmetric unimodal objective function assumption is in alignment with the operation of consistent estimators in the asymptotic region. The maximum likelihood estimator is the prime example for the consistent estimators. Specific to the maximum likelihood estimator, it has been shown that the suggested MSE expression reduces to the CRLB  and MCRLB in no-misspecification and misspecification cases, respectively. Furthermore, the suggested expression also yields the ZZB when an a-priori distribution is assigned to the unknown parameter for the MAP estimator.

An extension of the suggested MSE expression to the parameter estimation in the presence of nuisance parameters is given. Numerically friendly, but approximate, versions of the MSE expression are developed and some application examples are given. Numerical results show that the expression not only predicts the performance in the asymptotic region, but also provides valuable information in the threshold region. We consider that the applicability of the expression in other regions is related with the gradual degradation of asymptotic region operation conditions as the operating point moves from asymptotic region to the threshold region, say, with the reduction of SNR.

A possible interpretation for the MSE expression can be given in relation with the method of intervals (MIE). The MIE predicts the MSE by taking into account both small and gross error events via CRLB and the interval error probabilities, respectively. The suggested MSE expression for the ML estimator uses the likelihood ratio for the same purpose; but, it does not have a problem specific interval selection. 

Another interpretation for the MSE expression can be given in connection with the ZZB. As in ZZB, the suggested MSE expression is based on the pairwise error probabilities. Furthermore, the average of the expression for the MAP estimator exactly reproduces ZZB for random parameters. Hence, the suggested MSE expression for the ML estimator can also be considered, at least informally, as the non-random parameter version of the ZZB.

An interesting observation in the non-random parameter case was that,  for medium and low SNR, the proposed MSE expressions usually slightly underestimated the true MSE of the estimators. Hence, a potential future study is to investigate whether the proposed expressions have any lower bounding properties in medium and/or low SNR regions under some conditions.

\appendix
\section{Proof of Theorem~\ref{thm:IDEV}}
\label{app:proofofTheoremIDEbiasandMSE}
Note that the true estimator statistic $V_{\hat{\theta}}(\theta)$ defined in~\eqref{eqn:DefinitionofV} can be written as~\cite{BellSEVT:1997,KristenBell_PhDThesis}
\begin{align}
    V_{\hat{\theta}}(\theta)&=\,2\int_0^\infty\epsilon P\big(|\hat\theta-\theta|\ge \epsilon\big)\d\epsilon =\,2\int_0^\infty\epsilon \big[P\big(\hat\theta-\theta\ge \epsilon\big)+P\big(\hat\theta-\theta\le -\epsilon \big)\big]\d\epsilon,\label{eqn:MSEgeneral}
\end{align}
and the expression~\eqref{eqn:VsAreEqual} follows from~\eqref{eqn:MSEgeneral} if the equalities
\begin{subequations}
\label{eqn:equalitiestobeused}
\begin{align}
P(\hat\theta-\theta\ge \epsilon )=&\,P\left( \L(\mathbf{x};\theta+2\epsilon)\ge\L(\mathbf{x};\theta)\right),\label{eqn:tobeproven}\\
P(\hat\theta-\theta\le -\epsilon)=&\,P\left( \L(\mathbf{x};\theta-2\epsilon)\ge\L(\mathbf{x};\theta)\right),\label{eqn:nottobeproven}
\end{align}
\end{subequations}
hold for $\epsilon>0$. In the following we first show that the equalities in~\eqref{eqn:equalitiestobeused} indeed hold under the symmetry and unimodality assumptions of Theorem~\ref{thm:IDEV}. Only the proof of the equality $\eqref{eqn:tobeproven}$ will be made since the proof for~\eqref{eqn:nottobeproven} is very similar. In order to prove $\eqref{eqn:tobeproven}$, we will show that $\hat\theta-\theta\ge \epsilon$ if and only if $\L(\mathbf{x};\theta+2\epsilon)\ge \L(\mathbf{x};\theta)$. The proof has two parts.
\begin{itemize}
\item\textbf{Proof of the implication $\hat\theta-\theta\ge \epsilon$ $\Rightarrow$ $\L(\mathbf{x};\theta+2\epsilon)\ge \L(\mathbf{x};\theta)$:} Suppose that $\hat\theta\ge \theta+\epsilon$. Since $\epsilon> 0$, it is clear that $\theta<\hat\theta$. If $\theta+2\epsilon<\hat\theta$, since $\L(\mathbf{x};\theta)$ is strictly increasing for all $\theta<\hat{\theta}$ and since $\theta<\theta+2\epsilon< \hat{\theta}$, we would have $\L(\mathbf{x};\theta+2\epsilon)> \L(\mathbf{x};\theta)$ and this would make the inequality $\L(\mathbf{x};\theta+2\epsilon)\ge \L(\mathbf{x};\theta)$ hold. Hence, we only need to consider the case $\theta<\hat\theta\le \theta+2\epsilon$. In this case we will show that the inequality $\L(\mathbf{x};\theta+2\epsilon)\ge \L(\mathbf{x};\theta)$ holds by contraposition. Suppose that the reverse inequality, i.e., $\L(\mathbf{x};\theta+2\epsilon)< \L(\mathbf{x};\theta)$, holds. By the symmetry property we have
\begin{align}
    \L(\mathbf{x};\theta+2\epsilon)=&\,\L(\mathbf{x};\hat\theta+(\theta+2\epsilon-\hat\theta)) = \,\L(\mathbf{x};\hat\theta-(\theta+2\epsilon-\hat\theta))
    =\,\L(\mathbf{x};2\hat\theta-\theta-2\epsilon),\label{eqn:usingsymmetry}
    \end{align}
which shows that $\L(\mathbf{x};2\hat\theta-\theta-2\epsilon) < \L(\mathbf{x};\theta)$.
Since $2\hat\theta-\theta-2\epsilon\le\hat\theta$ (since $2\hat\theta-\theta-2\epsilon$ is the mirror image of $\theta+2\epsilon$ (with respect to $\hat\theta$), which is greater than or equal to $\hat\theta$) and since $\L(\mathbf{x};\theta)$ is strictly increasing for all $\theta<\hat{\theta}$, the inequality $\L(\mathbf{x};2\hat\theta-\theta-2\epsilon) < \L(\mathbf{x};\theta)$ implies that $2\hat\theta-\theta-2\epsilon<\theta$. This inequality is equivalent to the inequality $\hat\theta-\theta<\epsilon$, which completes the proof.

\item\textbf{Proof of the implication $\L(\mathbf{x};\theta+2\epsilon)\ge \L(\mathbf{x};\theta)$ $\Rightarrow$ $\hat\theta-\theta\ge \epsilon$:} Suppose that $\L(\mathbf{x};\theta+2\epsilon)\ge \L(\mathbf{x};\theta)$. Since $\L(\mathbf{x};\theta)$ is strictly decreasing for all $\theta>\hat{\theta}$, we cannot have $\theta>\hat{\theta}$. Hence, we need to have $\theta\le\hat\theta$. If $\theta+2\epsilon<\hat\theta$, we have $\hat\theta-\theta> 2\epsilon>\epsilon$, which makes the inequality $\hat\theta-\theta\ge \epsilon$ hold. Hence, we only need to consider the case $\theta\le\hat\theta\le \theta+2\epsilon$. By the symmetry property~\eqref{eqn:usingsymmetry} we see that $\L(\mathbf{x};2\hat\theta-\theta-2\epsilon)=\L(\mathbf{x};\theta+2\epsilon) \ge \L(\mathbf{x};\theta)$. Since we have $2\hat\theta-\theta-2\epsilon\le\hat\theta$ and $\theta\le\hat\theta$ and since  $\L(\mathbf{x};\theta)$ is increasing for all $\theta<\hat{\theta}$, we need to have $2\hat\theta-\theta-2\epsilon\ge \theta$. This inequality is equivalent to the inequality $\hat\theta-\theta\ge\epsilon$, which completes the proof.
\end{itemize}
Hence the equalities in~\eqref{eqn:equalitiestobeused} hold and we can write~\eqref{eqn:MSEgeneral} as
\begin{subequations}
    \begin{align}
        V_{\hat{\theta}}(\theta)\triangleq&\,2\int_0^\infty\epsilon \big[P\left(\L(\mathbf{x};\theta+2\epsilon)\ge\L(\mathbf{x};\theta) \right)+P\left(\L(\mathbf{x};\theta-2\epsilon)\ge\L(\mathbf{x};\theta) \right)\big]\d\epsilon,\\
        =&\,2\int_0^\infty\epsilon P\left( \L(\mathbf{x};\theta+2\epsilon)\ge\L(\mathbf{x};\theta) \right)\d\epsilon+2\int_0^\infty\epsilon P\left( \L(\mathbf{x};\theta-2\epsilon)\ge\L(\mathbf{x};\theta) \right)\d\epsilon,\\
        =&\,2\int_0^\infty\epsilon P\left( \L(\mathbf{x};\theta+2\epsilon)\ge\L(\mathbf{x};\theta) \right)\d\epsilon+2\int_0^{-\infty}\epsilon P\left( \L(\mathbf{x};\theta+2\epsilon)\ge\L(\mathbf{x};\theta) \right)\d\epsilon,\\
        =&\,2\int_0^\infty\epsilon P\left( \L(\mathbf{x};\theta+2\epsilon)\ge\L(\mathbf{x};\theta) \right)\d\epsilon-2\int^0_{-\infty}\epsilon P\left( \L(\mathbf{x};\theta+2\epsilon)\ge\L(\mathbf{x};\theta) \right)\d\epsilon,\\
        =&\,2\int^\infty_{-\infty}|\epsilon| P\left( \L(\mathbf{x};\theta+2\epsilon)\ge\L(\mathbf{x};\theta) \right)\d\epsilon\triangleq\widehat{V}_{\hat{\theta}}(\theta),
    \end{align}
\end{subequations}
which completes the proof.

\section{Proof of Proposition~\ref{prop:CaseofCRLB}}\label{app:ProofforCRLB}
We can write the probability in the integrand of
$\widehat\MSE_{\mathrm{ML}}(\bar\theta)$ in~\eqref{eqn:MSEHATMLdefinition} as
\begin{subequations}
    \begin{align}
        P\bigg( \myfrac[1.5pt]{f(\mathbf{x};\bar{\theta}+2\epsilon)}{f(\mathbf{x};\bar{\theta})}\ge 1\bigg)&=P\bigg(\ln\myfrac[1.5pt]{f(\mathbf{x};\bar{\theta}+2\epsilon)}{f(\mathbf{x};\bar{\theta})} \ge 0\bigg)=P\bigg(\frac{1}{N}\ln\myfrac[1.5pt]{f(\mathbf{x};\bar{\theta}+2\epsilon)}{f(\mathbf{x};\bar{\theta})}  \ge 0\bigg)\\
        &=P\bigg(\frac{1}{N}\ln\myfrac[1.5pt]{ f(\mathbf{x};\bar{\theta}+2\epsilon)}{f(\mathbf{x};\bar{\theta})} +D(\bar{\theta}||\bar{\theta}+2\epsilon)\ge D(\bar{\theta}||\bar{\theta}+2\epsilon)\bigg)\\
        &\le P\bigg(\bigg|\frac{1}{N}\ln\myfrac[1.5pt]{f(\mathbf{x};\bar{\theta}+2\epsilon)}{f(\mathbf{x};\bar{\theta})} +D(\bar{\theta}||\bar{\theta}+2\epsilon)\bigg| \ge D(\bar{\theta}||\bar{\theta}+2\epsilon)\bigg)\rightarrow 0
    \end{align}
\end{subequations}
for $\epsilon\neq 0$ as $N\rightarrow\infty$ where $D(\bar{\theta}||\bar{\theta}+2\epsilon)$ stands for $D(f(x;\bar{\theta})||f(x;\bar{\theta}+2\epsilon))$. This is because we have
\begin{align}
    \frac{1}{N}\ln\myfrac[1.5pt]{f(\mathbf{x}|\bar\theta+2\epsilon)}{f(\mathbf{x}|\bar\theta)}\overset{\text{p}}{\rightarrow} -D(\bar{\theta}||\bar{\theta}+2\epsilon)
\end{align}
as $N\rightarrow\infty$ by the law of large numbers and $D(\bar{\theta}||\bar{\theta}+2\epsilon)>0$ for $\epsilon\neq 0$ due to the assumption A3. As a result, as $N\rightarrow \infty$, the integration in~\eqref{eqn:MSEHATMLdefinition} will be effectively only over an infinitesimal neighborhood of $\epsilon=0$ and it is only the behavior of the probability $P\big( {f(\mathbf{x};\bar{\theta}+2\epsilon)}/{f(\mathbf{x};\bar{\theta})}\ge 1\big)$ as $\epsilon\rightarrow 0$ which determines the MSE expression $\widehat\MSE_{\mathrm{ML}}(\bar\theta)$ in~\eqref{eqn:MSEHATMLdefinition}.

Using the assumption A1, we can now obtain the Taylor expansion of $\ln f(\mathbf{x};\bar{\theta}+2\epsilon)$ around $\epsilon=0$ given as
\begin{align}
    \ln &f(\mathbf{x};\bar{\theta}+2\epsilon)=\ln f(\mathbf{x};\bar{\theta})+2\frac{\partial}{\partial\theta}\ln f(\mathbf{x};\bar{\theta})\epsilon+2\frac{\partial^2}{\partial\theta^2}\ln f(\mathbf{x};\bar{\theta})\epsilon^2+\frac{4}{3}\frac{\partial^3}{\partial\theta^3}\ln f(\mathbf{x};\tilde{\theta})\epsilon^3,
\end{align}
where $\tilde\theta$ is between $\bar\theta$ and $\bar\theta+2\epsilon$. Since the $\frac{\partial^3}{\partial\theta^3}\ln f(\mathbf{x};\tilde{\theta})$ is bounded by assumption A2 as $N\rightarrow\infty$, the approximation
\begin{align}
    \ln f(\mathbf{x};\bar{\theta}+2\epsilon)\approx&\ln f(\mathbf{x};\bar{\theta})+2\frac{\partial}{\partial\theta}\ln f(\mathbf{x};\bar{\theta})\epsilon+2\frac{\partial^2}{\partial\theta^2}\ln f(\mathbf{x};\bar{\theta})\epsilon^2
\end{align}
becomes valid as $\epsilon\rightarrow 0$. By rearranging, we can write
\begin{align}
 \ln \frac{f(\mathbf{x};\bar{\theta}+2\epsilon)}{f(\mathbf{x};\bar{\theta})} \approx 2\frac{\partial}{\partial\theta}\ln f(\mathbf{x};\bar{\theta})\epsilon+2\frac{\partial^2}{\partial\theta^2}\ln f(\mathbf{x};\bar{\theta})\epsilon^2\label{eqn:epsilon0loglikelihoodratio}
\end{align}
as $\epsilon\rightarrow 0$.

We can also write the Taylor expansion of $\frac{\partial}{\partial\theta}\ln f(\mathbf{x};\hat\theta)$ around $\theta=\bar\theta$ given as
\begin{align}
    0=\frac{\partial}{\partial\theta}\ln f(\mathbf{x};\hat\theta)=&\,\frac{\partial}{\partial\theta}\ln f(\mathbf{x};\bar\theta)+\frac{\partial^2}{\partial\theta^2}\ln f(\mathbf{x};\bar\theta)(\hat{\theta}-\bar{\theta})+\frac{1}{2}\frac{\partial^3}{\partial\theta^3}\ln f(\mathbf{x};\theta')(\hat{\theta}-\bar{\theta})^2,
\end{align}
where $\theta'$ is between $\hat{\theta}$ and $\bar{\theta}$.
Since the $\frac{\partial^3}{\partial\theta^3}\ln f(\mathbf{x};\theta)$ is bounded by assumption A2 as $N\rightarrow\infty$, the approximation
\begin{align}
    0\approx&\frac{\partial}{\partial\theta}\ln f(\mathbf{x};\bar\theta)+\frac{\partial^2}{\partial\theta^2}\ln f(\mathbf{x};\bar\theta)(\hat{\theta}-\bar{\theta})
\end{align}
becomes valid as $\hat{\theta}\overset{\text{a.s.}}{\rightarrow}\bar{\theta}$ as $N\rightarrow\infty$. Rearranging, we obtain
\begin{align}
    \frac{\partial}{\partial\theta}\ln f(\mathbf{x};\bar\theta)\approx -\frac{\partial^2}{\partial\theta^2}\ln f(\mathbf{x};\bar\theta)(\hat{\theta}-\bar{\theta})\label{eqn:largesamplescore}
\end{align}
as $N\rightarrow\infty$. Substituting $\frac{\partial}{\partial\theta}\ln f(\mathbf{x};\bar\theta)$ in~\eqref{eqn:largesamplescore} into~\eqref{eqn:epsilon0loglikelihoodratio}, we get
\begin{align}
 \ln \myfrac[1.5pt]{f(\mathbf{x};\bar{\theta}+2\epsilon)}{f(\mathbf{x};\bar{\theta})} \approx -2\epsilon\frac{\partial^2}{\partial\theta^2}\ln f(\mathbf{x};\bar\theta)(\hat{\theta}-\bar{\theta}-\epsilon)\label{eqn:epsilon0largesampleloglikelihoodratio}
\end{align}
as $\epsilon\rightarrow 0$ and $N\rightarrow\infty$. We can now substitute the result~\eqref{eqn:epsilon0largesampleloglikelihoodratio} into the the probability in the integrand of
$\widehat\MSE_{\mathrm{ML}}(\bar\theta)$ in~\eqref{eqn:MSEHATMLdefinition} to obtain
\begin{align}
    P&\bigg( \myfrac[1.5pt]{f(\mathbf{x};\bar{\theta}+2\epsilon)}{f(\mathbf{x};\bar{\theta})}\ge 1\bigg)= P\bigg(-\epsilon\frac{\partial^2}{\partial\theta^2}\ln f(\mathbf{x};\bar\theta)(\hat{\theta}-\bar{\theta}-\epsilon)\ge 0\bigg).\label{eqn:beforepositivityofFIM}
\end{align}
Using assumptions A1-A4, it can be shown that (See \cite[Lemma 2.1 Part-i]{Vuong1986} or~\cite[Lemma 4.1 Part-i]{FortunatiGG2017})
\begin{align}
    -\frac{1}{N}\frac{\partial^2}{\partial\theta^2}\ln f(\mathbf{x};\bar\theta)\overset{\text{p}}{\rightarrow}& -\E\bigg[\frac{\partial^2}{\partial\theta^2}\ln f(x;\bar\theta)\bigg] =\,\I(\bar{\theta})>0
\end{align}
as $N\rightarrow\infty$ where we used the law of large numbers. This allows us to write~\eqref{eqn:beforepositivityofFIM} as
\begin{align}
    P\bigg( \myfrac[1.5pt]{f(\mathbf{x};\bar{\theta}+2\epsilon)}{f(\mathbf{x};\bar{\theta})}\ge 1\bigg)=&\,P\big(\epsilon(\hat{\theta}-\bar{\theta}-\epsilon)\ge 0\big)
    =\begin{cases}
    P\big(\hat\theta-\bar{\theta}\ge \epsilon\big),&\epsilon>0\\
    1,&\epsilon=0\\
    P\big(\hat\theta-\bar{\theta}\le \epsilon\big),&\epsilon<0
    \end{cases},
    \label{eqn:afterpositivityofFIM}
\end{align}
for $\epsilon\rightarrow 0$ and $N\rightarrow\infty$. Note that the probabilities $P\big(\hat\theta-\bar{\theta}\ge \epsilon\big)$, $\epsilon>0$ and $P\big(\hat\theta-\bar{\theta}\le \epsilon\big)$, $\epsilon<0$ would vanish as $N\rightarrow\infty$, just as the probability $P\big( f(\mathbf{x};\bar{\theta}+2\epsilon)/f(\mathbf{x};\bar{\theta})\ge 1\big)$, $\epsilon\neq0$, itself, thanks to the fact that $\hat{\theta}\overset{\text{a.s.}}{\rightarrow}\bar\theta$.
As a result, we can substitute the right hand side of~\eqref{eqn:afterpositivityofFIM} into the finite support version of the integral~\eqref{eqn:MSEHATMLdefinition} to get
\begin{subequations}
\label{eqn:MLafterProbabilitySubstitution}
    \begin{align}
        \widehat{\MSE}_{\mathrm{ML}}(\bar\theta)&\triangleq 2\int_{\frac{\theta_{\mathrm{min}}-\bar\theta}{2}}^{\frac{\theta_{\mathrm{max}}-\bar\theta}{2}}|\epsilon|P\bigg( \myfrac[1.5pt]{f(\mathbf{x};\bar{\theta}+2\epsilon)}{f(\mathbf{x};\bar{\theta})}\ge 1\bigg)\d\epsilon
        =2\int_{\theta_{\mathrm{min}}-\bar\theta}^{\theta_{\mathrm{max}}-\bar\theta}|\epsilon|P\bigg( \myfrac[1.5pt]{f(\mathbf{x};\bar{\theta}+2\epsilon)}{f(\mathbf{x};\bar{\theta})}\ge 1\bigg)\d\epsilon\\
        &\hspace{-0.85cm}=-2\int_{\theta_{\mathrm{min}}-\bar\theta}^0\epsilon P\big(\hat\theta-\bar{\theta}\le \epsilon\big) \d\epsilon+2\int_0^{\theta_{\mathrm{max}}-\bar\theta}\epsilon P\big(\hat\theta-\bar{\theta}\ge \epsilon\big)\d\epsilon\\
        &\hspace{-0.85cm}=\int_{\theta_{\mathrm{min}}-\bar\theta}^0\epsilon^2 f_{\hat\theta-\bar\theta}(\epsilon) \d\epsilon+\int_0^{\theta_{\mathrm{max}}-\bar\theta} \epsilon^2 f_{\hat\theta-\bar\theta}(\epsilon)\d\epsilon=\int_{\theta_{\mathrm{min}}-\bar\theta}^{\theta_{\mathrm{max}}-\bar\theta} \epsilon^2 f_{\hat\theta-\bar\theta}(\epsilon)\d\epsilon =\E[(\hat\theta-\bar{\theta})^2]\triangleq\MSE_{\mathrm{ML}}(\bar{\theta}),
    \end{align}
\end{subequations}
 as $N\rightarrow\infty$, which completes the proof of~\eqref{eqn:ExpressionGoestoTrueMSE}. The proof of~\eqref{eqn:ExpressionGoestoCRLB} follows trivially if the ML estimate $\hat\theta$ is also asymptotically efficient.

\section{Proof of Proposition~\ref{prop:CaseofMCRLB}}\label{app:ProofforMCRLB}
In the case of MML estimation, we set $\L(\mathbf{x};\theta)\triangleq\ln f(\mathbf{x};\theta)$ in the probability in the integrand of
$\widehat{V}_{\hat\theta}(\theta)$ in~\eqref{eqn:VHATdefinition}. We can now write the probability in $\widehat{V}_{\hat\theta}(\theta)$ as
\begin{subequations}
    \begin{align}
        P\big(&\ln f(\mathbf{x};\theta_*+2\epsilon)\ge \ln f(\mathbf{x};\theta_*) \big)
        =P\bigg(\frac{1}{N}\ln\frac{f(\mathbf{x};\theta_*+2\epsilon)}{f(\mathbf{x};\theta_*)}  \ge 0\bigg)\\
        =&\,P\bigg(\frac{1}{N}\ln\frac{\bar{f}(\mathbf{x})}{f(\mathbf{x};\theta_*)}-\frac{1}{N}\ln\frac{\bar{f}(\mathbf{x})}{f(\mathbf{x};\theta_*+2\epsilon)}  \ge 0\bigg)\\
        =&\,P\bigg(\frac{1}{N}\ln\frac{\bar{f}(\mathbf{x})}{f(\mathbf{x};\theta_*)}-\frac{1}{N}\ln\frac{\bar{f}(\mathbf{x})}{f(\mathbf{x};\theta_*+2\epsilon)}-\big(D(\theta_*)-D(\theta_*+2\epsilon)\big)\ge D(\theta_*+2\epsilon)-D(\theta_*)\bigg)\\
        \le &\,P\bigg(\bigg|\frac{1}{N}\ln\frac{\bar{f}(\mathbf{x})}{f(\mathbf{x};\theta_*)}-\frac{1}{N}\ln\frac{\bar{f}(\mathbf{x})}{f(\mathbf{x};\theta_*+2\epsilon)}-\big(D(\theta_*)-D(\theta_*+2\epsilon)\big)\bigg|\ge D(\theta_*+2\epsilon)-D(\theta_*)\bigg)\rightarrow 0
    \end{align}
\end{subequations}
for $\epsilon\neq 0$ as $N\rightarrow\infty$ where $D(\theta)$ stands for $D(\bar{f}(x)||f(x;\theta))$. This is because we have
\begin{align}
\frac{1}{N}\ln\frac{\bar{f}(\mathbf{x})}{f(\mathbf{x};\theta_*)}-\frac{1}{N}\ln\frac{\bar{f}(\mathbf{x})}{f(\mathbf{x};\theta_*+2\epsilon)}\overset{\text{p}}{\rightarrow} D(\theta_*)-D(\theta_*+2\epsilon)
\end{align}
as $N\rightarrow\infty$ by the law of large numbers and $D(\theta_*+2\epsilon)>D(\theta_*)$ for $\epsilon\neq 0$ due to the assumption A3. As a result, as $N\rightarrow \infty$, the integration in~\eqref{eqn:MSEHATdefinition} will be effectively only over an infinitesimal neighborhood of $\epsilon=0$ and it is only the behavior of the probability $P\big( \ln f(\mathbf{x};\theta_*+2\epsilon)\ge \ln f(\mathbf{x};\theta_*)\big)$ as $\epsilon\rightarrow 0$ which determines the expression $\widehat{V}_{\hat\theta}(\theta_*)$.

Following a similar approach that is used for obtaining~\eqref{eqn:epsilon0loglikelihoodratio}, we can write
\begin{align}
 \ln \frac{f(\mathbf{x};\theta_*+2\epsilon)}{f(\mathbf{x};\theta_*)} \approx 2\frac{\partial}{\partial\theta}\ln f(\mathbf{x};\theta_*)\epsilon+2\frac{\partial^2}{\partial\theta^2}\ln f(\mathbf{x};\theta_*)\epsilon^2\label{eqn:epsilon0loglikelihoodratioMCRB}
\end{align}
as $\epsilon\rightarrow 0$. Using an approach similar to that used for obtaining~\eqref{eqn:largesamplescore} we can get
\begin{align}
    \frac{\partial}{\partial\theta}\ln f(\mathbf{x};\theta_*)\approx -\frac{\partial^2}{\partial\theta^2}\ln f(\mathbf{x};\theta_*)(\hat{\theta}-\theta_*)\label{eqn:largesamplescoreMCRB}
\end{align}
as $N\rightarrow\infty$. Substituting $\frac{\partial}{\partial\theta}\ln f(\mathbf{x};\theta_*)$ in~\eqref{eqn:largesamplescoreMCRB} into~\eqref{eqn:epsilon0loglikelihoodratioMCRB}, we get
\begin{align}
 \ln \frac{f(\mathbf{x};\theta_*+2\epsilon)}{f(\mathbf{x};\theta_*)} \approx -2\epsilon\frac{\partial^2}{\partial\theta^2}\ln f(\mathbf{x};\theta_*)(\hat{\theta}-\theta_*-\epsilon)\label{eqn:epsilon0largesampleloglikelihoodratioMCRB}
\end{align}
as $\epsilon\rightarrow 0$ and $N\rightarrow\infty$. We can now substitute the result~\eqref{eqn:epsilon0largesampleloglikelihoodratioMCRB} into the probability in the integrand of $\widehat{V}(\theta_*)$
\begin{align}
    P&\big(\ln f(\mathbf{x};\theta_*+2\epsilon)\ge \ln f(\mathbf{x};\theta_*)\big)= P\bigg(-\epsilon\frac{\partial^2}{\partial\theta^2}\ln f(\mathbf{x};\theta_*)(\hat{\theta}-\theta_*-\epsilon)\ge 0\bigg).\label{eqn:beforepositivityofFIMMCRB}
\end{align}
Using the assumptions A1-A4, it can be shown that (See \cite[Lemma 2.1 Part-i]{Vuong1986} or~\cite[Lemma 4.1 Part-i]{FortunatiGG2017})
\begin{align}
    -\frac{1}{N}\frac{\partial^2}{\partial\theta^2}\ln f(\mathbf{x};\theta_*)\overset{\text{p}}{\rightarrow}& -\E_{\bar{f}}\bigg[\frac{\partial^2}{\partial\theta^2}\ln f(x;\theta_*)\bigg]=-\mathcal{A}(\theta_*)>0
\end{align}
as $N\rightarrow\infty$ where we used the law of large numbers. This allows us to write~\eqref{eqn:beforepositivityofFIMMCRB} as
\begin{align}
    P\big(\ln f(\mathbf{x};\theta_*+2\epsilon)\ge&\,\ln f(\mathbf{x};\theta_*)\big)=\,P\big(\epsilon(\hat{\theta}-\theta_*-\epsilon)\ge 0\big)
    =\begin{cases}
        P\big(\hat\theta-\theta_*\ge \epsilon\big),&\epsilon>0\\
        1,&\epsilon=0\\
        P\big(\hat\theta-\theta_*\le \epsilon\big),&\epsilon<0
     \end{cases},
    \label{eqn:afterpositivityofFIMMCRB}
\end{align}
for $\epsilon\rightarrow 0$ and $N\rightarrow\infty$. Note that the probabilities $P\big(\hat\theta-\theta_*\ge \epsilon\big)$, $\epsilon>0$ and $P\big(\hat\theta-\theta_*\le \epsilon\big)$, $\epsilon<0$ would vanish as $N\rightarrow\infty$, just as the probability $P\big( \ln f(\mathbf{x};\theta_*+2\epsilon)\ge \ln f(\mathbf{x};\theta_*)\big)$, $\epsilon\neq0$, itself, thanks to the fact that $\hat{\theta}\overset{\text{a.s.}}{\rightarrow}\theta_*$.
As a result, we can substitute the right hand side of~\eqref{eqn:afterpositivityofFIMMCRB} into the finite support version of the integral in $\widehat{V}_{\hat\theta}(\theta_*)$ to get
\begin{subequations}
\label{eqn:MLafterProbabilitySubstitutionMCRB}
\begin{align}
\widehat{V}_{\text{MML}}(\theta_*)\triangleq& 2\int_{\frac{\theta_{\mathrm{min}}-\theta_*}{2}}^{\frac{\theta_{\mathrm{max}}-\theta_*}{2}}|\epsilon|P\left( \ln f(\mathbf{x};\theta_*+2\epsilon)\ge \ln f(\mathbf{x};\theta_*)\right)\d\epsilon\\
&= 2\int_{\theta_{\mathrm{min}}-\theta_*}^{\theta_{\mathrm{max}}-\theta_*}|\epsilon|P\left( \ln f(\mathbf{x};\theta_*+2\epsilon)\ge \ln f(\mathbf{x};\theta_*)\right)\d\epsilon\\
&=-2\int_{\theta_{\mathrm{min}}-\theta_*}^0\epsilon P\big(\hat\theta-\theta_*\le \epsilon\big) \d\epsilon+2\int_0^{\theta_{\mathrm{max}}-\theta_*}\epsilon P\big(\hat\theta-\theta_*\ge \epsilon\big)\d\epsilon\\
&=\int_{\theta_{\mathrm{min}}-\theta_*}^0\epsilon^2f_{\hat\theta-\theta_*}(\epsilon)\d\epsilon+\int_0^{\theta_{\mathrm{max}}-\theta_*}\epsilon^2 f_{\hat\theta-\theta_*}(\epsilon)\d\epsilon\\
&=\int_{\theta_{\mathrm{min}}-\theta_*}^{\theta_{\mathrm{max}}-\theta_*} \epsilon^2 f_{\hat\theta-\bar\theta}(\epsilon)\d\epsilon\\
&=\E[(\hat\theta-\theta_*)^2]\triangleq V_{\text{MML}}(\theta_*),
\end{align}
\end{subequations}
 as $N\rightarrow\infty$, which completes the proof of~\eqref{eqn:ExpressionGoestoTrueV}. The proof of~\eqref{eqn:ExpressionGoestoMCRLB} follows trivially if the MML estimate $\hat\theta$ is also asymptotically misspecified efficient.

\section{Implementation Details of the Methods Used in Section~\ref{sec:numericalResults}}\label{app:implementationDetails}

\subsection{Implementation Details for Section~\ref{sec:DOAest}}
        \begin{itemize}
            \item ML estimate is calculated on a grid. The grid consists of 3600 uniformly spaced points in the interval $[-\pi,\,\pi]$ for (scalar) azimuth estimation, and 3600 uniformly spaced points in the interval $[0,\,\pi]$ for (scalar) elevation estimation. For the case of nuisance parameter, azimuth and elevation angles are selected such that the corresponding unit vectors have an approximately uniform distribution over the unit sphere. To do so, elevation interval $[0,\,\pi]$ is divided into 200 equally spaced values $\{ \theta_k\}_{k=1}^{200}$, and each constant elevation circle on the unit sphere is divided into $\lceil{(100\sin(\theta_k)}\rceil$ points where $\lceil{\cdot}\rceil$ denotes the ceiling function. By doing so, a total of $198192$ grid points ($\phi$-$\theta$ pairs) are obtained. 
            
            BBs~\cite{BarankinBound1971} are calculated with a single test point which is optimized on a grid. The single and multiple parameter BB grids are selected the same as the grids used for ML.
            MIE grid points are selected as the local maxima outside the mainlobe in the beampattern for the scalar parameter estimation case, which requires a peak finding algorithm over the 1D beampattern function. For the case of a nuisance parameter, the beampattern function is a 2D surface, and grid points are the local maxima on this surface. For both cases the beampattern function is calculated over the ML grid, and local maxima are found using Matlab built-in functions (\texttt{findpeaks($\cdot$)} for maxima on the 1D curve, \texttt{imregionalmax($\cdot$)} for maxima on the 2D surface).
            
            Proposed method does not require grid point selection for the scalar parameter case. For the case of a nuisance parameter, grid points are selected over the entire support of the nuisance parameter with logarithmic spacing around the true value, making the grid denser as the grid points approach the true parameter value. In Matlab notation, the grid is defined as follows
            \begin{subequations}
            \begin{align}
                \boldsymbol{\delta\theta} &=  \texttt{logspace}(-7, \mathrm{log_{10}}e_{\text{max}}, 60), \label{eq:logspace} \\
                \boldsymbol{\Theta}_{\backslash 1} &= [\bar{\theta},\, \bar{\theta} - \boldsymbol{\delta\theta},\, \bar{\theta} + \boldsymbol{\delta\theta}].
            \end{align}
            \end{subequations}
            The statement~\eqref{eq:logspace} generates 60 logarithmically spaced points within $[10^{-7},e_{\text{max}} ]$ interval in Matlab. We selected $e_{\text{max}} =\pi/2$ for the elevation angle as the nuisance parameter and $e_{\text{max}} =\pi$ for the azimuth angle as the nuisance parameter.  
        \end{itemize}
\subsection{Implementation Details for Section~\ref{sec:DOAest_modelMismatch}}
        \begin{itemize}
            \item ML grid consists of 3600 uniformly spaced points in the interval $[-\pi,\,\pi]$ for (scalar) azimuth estimation. BB grid is the same as the ML grid. Other methods do not require grid points.
        \end{itemize}
\subsection{Implementation Details for Section~\ref{IDEexample}}
        \begin{itemize}
        \item BB grid is the same as the ML grid for the problem in Section~\ref{sec:DOAest_modelMismatch}. Other methods do not require grid points.
\end{itemize}
The numerical integrals of the proposed method in all subsections up to this point are taken using the Matlab function  \texttt{integral($\cdot$)} with the following tolerance values: \texttt{AbsTol=1e-5}, \texttt{RelTol=1e-5}. The integration limits for the numerical integrals of the proposed method were set as $\epsilon \in \big[\frac{-\pi-\bar{\phi}}{2}, \frac{\pi-\bar{\phi}}{2}\big]$ for azimuth estimation and $\epsilon \in \big[-\frac{\bar{\theta}}{2}, \frac{\pi-\bar{\theta}}{2}\big]$ for elevation estimation.
\subsection{Implementation Details for Section~\ref{sec:bayesianDOA}}
        \begin{itemize} 
        \item ML and MAP use the same grid, which consists of 8192 uniformly spaced points over $[-\pi, \pi]$ for $\omega = \pi\cos(\phi)$. The (non-uniform) grid points for $\phi$ are calculated using expression $\arccos(\frac{\omega}{\pi})$ from the grid for~$\omega$.
        \item The double integral for ZZB is taken using the MATLAB function \texttt{integral2($\cdot$)} with the default tolerance settings. 
        \item The numerical integrals of $\widehat{\MSE}_{\mathrm{ML}}(\phi)$ and $\widehat{\MSE}_{\mathrm{MAP}}(\phi)$ are taken using the Matlab function  \texttt{integral($\cdot$)} with the  tolerance values \texttt{AbsTol=1e-18}, \texttt{RelTol=1e-12}. The integrals with respect to the prior are calculated over a uniform grid over the inverval $[0,\pi]$ with the grid spacing 0.01.
\end{itemize}


 \bibliographystyle{elsarticle-num}
 \setlength{\bibsep}{-1pt plus 0ex}
 \bibliography{references}





\end{document}